\documentclass[11pt,reqno]{amsart}
\usepackage{amsfonts,amsmath}
\usepackage{graphicx}
\usepackage{epstopdf}

\vspace{4ex}
\begin{document}
\vspace*{1cm}
\begin{center}
{\Large \bf Deformed quantum Calogero-Moser problems and Lie
superalgebras}
\\*[4ex]
{\bf A.N. Sergeev$^{\dagger,\ddagger}$ and
A.P. Veselov$^{\ddagger,\star}$} \\*[2ex]
\end{center}

\noindent $^\dagger$ Balakovo Institute of Technology and
Control, Balakovo,413800, Russia

\noindent $^\ddagger$
Department of Mathematical Sciences, Loughborough University,\\
Loughborough,  LE11 3TU, UK

\noindent $^\star$ Landau Institute
for Theoretical Physics, Kosygina 2, Moscow, 117940, Russia

\noindent E-mail addresses: A.N. Sergeev@lboro.ac.uk,
A.P.Veselov@lboro.ac.uk

\vspace*{1cm} %
\vspace*{1cm}

{\small  {\bf Abstract.} The deformed quantum
Calogero-Moser-Sutherland problems related to the root systems of
the contragredient Lie superalgebras are introduced. The
construction is based on the notion of the generalized root
systems suggested by V. Serganova. For the classical series a
recurrent formula for the quantum integrals is found, which
implies the integrability of these problems. The corresponding
algebras of the quantum integrals are investigated, the explicit
formulas for their Poincare series for generic values of the
deformation parameter are presented.}

\section{Introduction}
Calogero-Moser-Sutherland (CMS) problem in its original form
\cite{C},\cite{S} describes the particles on the line pairwise
interacting  with the potential $$ U(x_{1},\dots ,
x_{n})=\sum_{1\le i<j\le n}\frac{g^2\omega ^2}{\sin
^2\omega(x_{i}-x_{j})} $$ In the limit $\omega\to 0$ one has the
rational potential $$ U(x_{1},\dots , x_{n})=\sum_{1\le i<j\le
n}\frac{g^2}{(x_{i}-x_{j})^2} $$ Olshanetsky and Perelomov
\cite{OP} proposed a generalization of these problems related to
any root system. Corresponding quantum Hamiltonian has a form
\begin{equation}
\label{CMS}
L=-\Delta + \sum_{\alpha\in R_{+}}\frac{m_{\alpha}(m_{\alpha}+2m_{2\alpha}+1)(\alpha,\alpha)}{\sin^2(\alpha,x)}
\end{equation}
where $R_{+}$ is a positive part of a root system $R$ (which could
be non-reduced), and $m(\alpha)=m_{\alpha}$ is a function on $R$
which is invariant under the corresponding Weyl group $W$. If $R$
is a root system of a compact symmetric space $X$ and
$m_{\alpha}=\frac{\mu_{\alpha}}{2}-1$ where $\mu_{\alpha}$ are the
multiplicities of the roots then the operator (\ref{CMS}) is
conjugated to the radial part of the Laplace-Beltrami operator on
$X$
\begin{equation}
\label{rad} {\mathcal L} =-\Delta + 2\sum_{\alpha\in
R_{+}}m_{\alpha}\cot(\alpha,x)\partial_{\alpha},
\end{equation}
namely $$ {\mathcal L}={\hat\psi}_{0}^{-1}\circ L\circ
{\hat\psi}_{0} $$ where ${\hat\psi}_{0}$ is the multiplication
operator by the function
\begin{equation}
\label{psi0}
{\hat\psi}_{0}=\prod_{\alpha\in R_{+}} \sin^{-m_{\alpha}}(\alpha,x).
\end{equation}

In this general form this was shown by Olshanetsky and Perelomov
\cite{OP2} but in the special case $X=SU(n)/SO(n)$ it was observed
already in 1964 by Berezin et al in \cite{BPF}. So integrability
of the generalized CMS quantum problem (\ref{CMS}) in this case
follows from the general theory of the invariant operators on the
symmetric spaces (see e.g. \cite{Hel}). In the general case the
proof of integrability was found later by Heckman and Opdam
\cite{HO,H} who used very different arguments.

It turned out however that there are other, non-symmetric
integrable generalizations of the quantum CMS problems. The first
series $A_n(m)$ of such "deformed" CMS problems have been found by
Chalykh, Feigin and one of the authors in \cite{CFV},\cite{CFV1},
later another series $C_n(m,l)$ was discovered \cite{CFV2}.
Although these deformations since then appeared in different
context (see e.g. \cite{V1} about relations to WDVV equations) the
algebraic nature of them until recently was totally unclear.

An important step was done by one of the authors in
\cite{Ser,Ser1} who observed that the deformed CMS operator of the
type $A_{n}(m)$ for a special value of parameter can be
interpreted as the radial part of Laplace-Beltrami operator on
certain symmetric superspace. This indicated that the clue to this
mystery could be found in this direction.

The goal of the present paper is to develop a systematic theory of
the deformed CMS quantum systems related to Lie superalgebras and
symmetric superspaces. A crucial role in our approach plays the
notion of the {\it generalized root systems} introduced by V.
Serganova in the paper \cite{Serga}.

Serganova suggested a version of the standard geometric
description of the root systems \cite{Bourbaki} in the presence of
the isotropic roots. She classified all such irreducible systems
and showed that they are essentially the root systems of the
contragredient Lie superalgebras classified by V. Kac in
\cite{Kac}. All the generalized root systems have partial symmetry
described by the Weyl group $W_{0}$ generated by reflections
corresponding to the non-isotropic ("real") roots.

For any such generalized root system $R$ we construct a family of
the deformed quantum CMS systems in the following way. The system
$R$ stays the same, we change the scalar product and the
multiplicities of the roots (which are not integers anymore) in
such a way that the following conditions are satisfied:

1) the new bilinear form $B$ and the multiplicity function stay $W_{0}$-invariant

2) multiplicities of all the imaginary roots are fixed to be 1

3) the corresponding Schr\"odinger operator (\ref{CMS}) has a radial form (\ref{rad}) or equivalently, has an eigenfunction of the form (\ref{psi0}).

The last condition leads to certain relations between the
multiplicities and parameters of the form. Our analysis shows that
the admissible forms for all generalized root systems depend on
one deformation parameter (in the case $D(2,1,\lambda)$ we have
three parameters but two of them are already in the Lie
superalbegra itself), although multiplicities have sometimes more
freedom (see Section 2). We call the corresponding operators the
{\it deformed CMS operators related to generalized root system}
$R$.

According to Serganova's classification we have two infinite
series of such operators related to the  classical systems $R$ of
the type $ A(n,m)$ and $BC(n,m)$, and three exceptional cases
corresponding to the exceptional systems $G(1,2) $, $AB(1,3)$ and
$D(2,1,\lambda)$. The deformed operators corresponding to the
classical series have the form:
\begin{eqnarray}
\label{anm} L_{A(n-1,m-1)}& = &-\left (\frac{\partial^2}{{\partial
x_{1}}^2}+\dots +\frac{\partial^2}{{\partial x_{n}}^2}\right)-
k\left(\frac{\partial^2}{{\partial y_{1}}^2}+\dots
+\frac{\partial^2}{{\partial y_{m}}^2}\right)
+\sum_{i<j}^{n}\frac{2k(k+1)}{\sin^2(x_{i}-x_{j})} \nonumber \\ &
&+\sum_{i<j}^{m}\frac{2(k^{-1}+1)}{\sin^2(y_{i}-y_{j})}
+\sum_{i=1}^{n}\sum_{j=1}^{m}\frac{2(k+1)}{\sin^2(x_{i}-y_{j})}
\end{eqnarray}
for $ A(n-1,m-1)$, where $k$ is an arbitrary parameter, and
\begin{eqnarray}
\label{bcnm} L_{BC(n,m)}& =& -\left(\frac{\partial^2}{{\partial
x_{1}}^2}+\dots +\frac{\partial^2}{{\partial x_{n}}^2}\right)
-k\left(\frac{\partial^2}{{\partial y_{1}}^2}+ \dots
+\frac{\partial^2}{{\partial y_{m}}^2} \right) \nonumber \\& &
+\sum_{i<j}^{n}\left(\frac{2k(k+1)}{\sin^2(x_{i}-x_{j})}+\frac{2k(k+1)}{\sin^2(x_{i}+x_{j})}\right)
+\sum_{i<j}^{m}\left(\frac{2(k^{-1}+1)}{\sin^2(y_{i}-y_{j})}+\frac{2(k^{-1}+1)}{\sin^2(y_{i}+y_{j})}\right)
\nonumber
\\& & +\sum_{i=1}^{n}\sum_{j=1}^{m}\left(\frac{2(k+1)}{\sin^2(x_{i}-y_{j})}+
\frac{2(k+1)}{\sin^2(x_{i}+y_{j})}\right) +\sum_{i=1}^n
\frac{p(p+2q+1)}{\sin^2x_{i}} \nonumber \\& & +\sum_{i=1}^n
\frac{4q(q+1)}{\sin^22x_{i}} +\sum_{j=1}^m \frac{k
r(r+2s+1)}{\sin^2y_{j}}+\sum_{j=1}^m \frac{4k
s(s+1)}{\sin^22y_{j}}
\end{eqnarray}
for $BC(n,m)$, where the parameters $k,p,q,r,s$ satisfy the
following relations
\begin{equation}
\label{rel}
p=kr,\quad 2q+1=k(2s+1)
\end{equation}

The system (\ref{anm}) can be considered as the interaction of two
groups of particles of masses 1 and $\frac{1}{k}$ respectively
with the special parameters of interaction depending on $k.$ When
$m=1$ ( i.e. when the second group consists only of one particle)
this system was first proposed in \cite{CFV}. For the general $n$
and $m$ the operator (\ref{anm}) was first introduced by one of
the authors in \cite{Ser} but its rational limit was discovered
earlier by Berest and Yakimov, who were looking for a Darboux-type
transformations for Calogero-Moser systems \cite{BY}.

The system (\ref{bcnm}) can be interpreted in a similar way under
the assumption that the configuration of particles is symmetric
with respect to the origin. Although it depends on 5 parameters
only 3 of them are independent due to relations (\ref{rel}) (say
$k,p$ and $q$). The system $BC(n,m)$ with $m=1$ and $p=0$ was
first considered in \cite{CFV2}. The case $m=1$ is special as the
only one when all the parameters could be integer. As far as we
know the operator (\ref{bcnm}) for the general $m,n$ as well as
the deformed CMS systems related to the exceptional root systems
$G(1,2), AB(1,3)$ and $D(2,1,\lambda)$ (see the next section) were
not considered before.

The conjecture is that all these quantum systems are integrable in
the sense that they have enough commuting integrals. In this paper
we prove this conjecture for the classical series ( i.e. for the
operators (\ref{anm})  and (\ref{bcnm})) explicitly constructing
the integrals.

The corresponding algebra of integrals seems to be interesting by
itself. To describe it we introduce the following algebra
$\Lambda_{R,B}$ related to a deformed generalized root system
$(R,B).$ It consists of the polynomials $p(x)$ on $V$ which are
invariant under reflections $s_{\alpha}$ corresponding to the real
roots (i.e. are $W_0$-invariant) and satisfy the conditions
\begin{equation}
\label{lambda}
 p(x+\frac{1}{2}\alpha)\equiv p(x-\frac{1}{2}\alpha)
\end{equation}
on the hyperplane $B(\alpha,x)=0$ for each imaginary root
$\alpha.$

Our main result can be formulated as follows.

\medskip
{\bf Theorem.} {\it For the classical generalized root systems $R$
and generic values of the deformation parameter $k$ in the form
$B$ there exists a monomorphism $\chi$ from the commutative
algebra $\Lambda_{R,B}$ into the algebra of differential operators
on $V$ such that $\chi(x^2)$ is the corresponding deformed CMS
operator related to $R$.}
\medskip

In the rational limit a similar result holds for a closely related
algebra $\Lambda^0_{R,B}$ with the condition (\ref{lambda}) in the
definition of $\Lambda_{R,B}$ replaced by its differential
version: $\partial_{\alpha} p(x) = 0$ when $B(\alpha,x)=0.$

The structure of the paper is following. We start with the precise
definition of the generalized root systems $R$ and formulate
Serganova's classification theorem. Then we define the deformed
CMS operators related to such a system $R$ and classify all of
them for each generalized root system.

In section 3 we prove the integrability of the deformed quantum
CMS problems for the classical series $A(n,m)$ and $BC(n,m)$. The
proof is effective: the quantum integrals are given explicitly by
some recurrent formula. Our formula can be considered as a
deformed version of the Matsuo's formula (2.3.6) from
\cite{Matsuo}.

In section 4 we introduce and investigate the algebras
$\Lambda_{R,B}$ and $\Lambda^0_{R,B}.$ In particular we show that
for the classical series and generic values of the deformation
parameter $k$ these algebras are finitely generated and compute
the Poincare series for them. We show that the image of the
Harish-Chandra homomorphism from the rings of quantum integrals of
the deformed CMS problems described in section 2 for generic $k$
is exactly the algebra $\Lambda_{R,B}$ (in the rational case -
$\Lambda^0_{R,B}$). In the last section 5 we discuss the elliptic
and difference generalizations of our operators.

\section{Generalized root systems and deformed quantum CMS problems.}

We start with the definition of the generalized root systems due
to V. Serganova \cite{Serga}. We should mention that there are
three slightly different definitions of generalized root systems
in \cite{Serga}, we choose one of them which suits best for our
purpose.

Let $V$ be a finite dimensional complex vector space with a
non-degenerate bilinear form $< , >$.

{\bf Definition.} The finite set $R\subset V\setminus\{0\}$ is
called a {\it generalized root system} if the following conditions
are fulfilled :

1) $R$ spans $V$ and $R=-R$ ;

2) if $\alpha,\beta\in R$ and  $<\alpha ,\alpha >\ne 0$ then
$\frac{2<\alpha ,\beta >}{<\alpha ,\alpha >}\in {\bf Z}$ and
$s_{\alpha}(\beta)=\beta -\frac{2<\alpha ,\beta >}{<\alpha ,\alpha
>}\alpha\in R$;

3) if $\alpha\in R$ and $<\alpha ,\alpha >=0$ then for any
$\beta\in R$ such that $<\alpha ,\beta >\ne 0$ at least one of the
vectors $\beta + \alpha$ or $\beta - \alpha$ belongs to $R$.

The non-isotropic roots are called {\it real}, the isotropic roots are
called {\it imaginary:}
$$
R_{re}=\{\alpha\in R: <\alpha,\alpha>\ne 0\}\quad
R_{im}=\{\alpha\in R: <\alpha,\alpha>= 0\}
$$

A generalized root system $R$ is called {\it reducible} if it can
be represented as a direct sum of two non-empty generalized root
systems $R_{1}$ and
 $R_{2}$, i.e. $V=V_{1}\oplus V_{2}$, $R_{1}\subset V_{1}$, $R_{2}\subset V_{2}$,
and $R=R_{1}\cup R_{2}$. Otherwise the system is called {\it irreducible.}

Any generalized root system has a partial symmetry described by
the finite group $W_0$ generated by the reflections with respect
to the real roots.

The main result of a remarkable Serganova's paper \cite{Serga} is
the classification theorem for the irreducible generalized root
systems which says that they all are contained in the following
list.

\medskip
{\bf List of the irreducible generalized root systems.}
\medskip

{\bf Classical series}
\medskip

{\bf 1. $A(n-1,m-1), \quad n \ne m$}

Let $V_{n,m} = V_{1}\oplus V_{2}$ be a vector space with the basis
$\{ e_{1},\dots , e_{n+m}\}$, such that $\{ e_{1},\dots , e_{n}\}$
be a basis of $V_{1}$ and $\{ e_{n+1},\dots , e_{n+m}\}$ be a
basis of $V_{2}$. Let $e^i, i = 1,..., n+m$ denote the
corresponding basis in the dual space  $V_{n,m}^*$.

Consider the following bilinear (indefinite) symmetric form on
$V_{n,m}$
\begin{equation}
\label{B} B(u,v) = \sum_{i=1}^n u^iv^i - \sum_{j=n+1}^{n+m}
u^jv^j,
\end{equation}
where $u^i, v^i$ are the coordinates of vectors $u,v$ in the basis
$e_i.$

Let us split the set of the indices $I = \{1, \dots, n+m\}$ into
two groups: $I=I_{1}\cup I_{2}$, where $I_{1}=\{1, \dots ,n\}$,
$I_{2}=\{n+1, \dots , n+m\}$ and rewrite the last formula as
\begin{equation}
\label{B} B = \sum_{i \in I_1} e^i\otimes e^i - \sum_{j \in I_2}
e^j\otimes e^j,
\end{equation}
where $B$ is now considered as an element of $V^* \otimes V^*.$

The generalized root system of type $A(n-1,m-1)$, $n\ne m$ is
defined as the set $R = {\{e_{i}-e_{j}, i\ne j, i,j\in I\}}$ and
the corresponding space $V$ is the hyperplane in $V_{n,m}$
generated by this set with the induced bilinear form. It is easy
to see that in this case $R_{re}=A_{n-1}\oplus A_{m-1}$, $R_{im}=
\{\pm(e_{i}-e_{j}), i\in I_{1}, j\in I_{2} \}$ Corresponding Lie
superalgebra is $sl(n|m).$

{\bf 2. $A(n-1,n-1)$}

If we would do the same when $m=n$ then we will have a problem:
the restriction of the form $B$ on the corresponding hyperplane
$V$ is degenerate. Indeed the vector $v=\sum_{i\in I_{1}} e_{i} -
\sum_{j\in I_{2}} e_{j}$ belongs to $V$ and orthogonal to all the
roots (and thus to all $V$). In order to have a proper generalized
root system in this case we should consider the quotient $V'=
V/<v>$ and the corresponding  set $R'$ which is the image of $R$
after such a projection. This is the system of the type
$A(n-1,n-1)$. Corresponding Lie superalgebra is $psl(n|n).$

{\bf 3. $B(n,m)$}

In this case $V=V_{n,m}$ which is defined above with the same bilinear form $B$
and $R$ is the set $\{\pm e_{i}\pm e_{j}, i\ne j, i,j\in I, \pm
e_{i}, i\in  I, \pm 2e_{i}, i\in I_{2} \}$. The real and imaginary roots are
$R_{re}=B_{n}\oplus BC_{m}$, $R_{im}=
\{\pm e_{i}\pm e_{j}, i\in I_{1}, j\in I_{2}
\}.$
This corresponds to the orthosymplectic Lie superalgebra $osp(2n+1|2m)$.

 {\bf 4. $D(n,m)$, $n \geq 2$}

$V=V_{n,m}$ is the same as in the previous case, but $R$ is the
set $\{\pm e_{i}\pm e_{j}, i\ne j, i,j\in I, \pm 2e_{i}, i\in
I_{2} \}$. We have $R_{re}=D_{n}\oplus C_{m}$, $R_{im}= \{\pm
e_{i}\pm e_{j}, i\in I_{1}, j\in I_{2} \}$ Corresponding Lie
superalgebra is $osp(2n|2m).$

{\bf 5. $C(0,m)$}

Here $V=V_{1,m}$ and
$R$ is the set $\{\pm e_{i}\pm e_{j}, i\ne j, i,j\in I, \pm
2e_{i}, i\in I_{2} \}$. In this case
$R_{re}= C_{m}$, $R_{im}= \{\pm e_{1}\pm e_{j}, j\in I_{2},\}.$
Corresponding Lie superalgebra is $osp(2|2m).$

{\bf 6. $C(n,m)$}

Here $V= V_{n,m}$ and
$R$ is the set $\{\pm e_{i}\pm e_{j}, i\ne j, i,j\in I,  \pm
2e_{i}, i\in I \}$, so that
$R_{re}=C_{n}\oplus C_{m}$, $R_{im}= \{\pm e_{i}\pm e_{j}, i\in
I_{1}, j\in I_{2}, \}.$
In this case and in the next one there are no related Lie superalgebras but there
are symmetric superspaces with such root systems.

 {\bf 7. $BC(n,m)$}

$V=V_{n,m}$ and $R$ consists of $\{\pm e_{i}\pm e_{j}, i\ne j, i,j\in I, \pm
e_{i},  \pm 2e_{i}, i\in I \}$. In this case $R_{re}=BC_{n}\oplus BC_{m}$, $R_{im}= \{\pm e_{i}\pm e_{j}, i\in I_{1}, j\in I_2 \}$

\medskip

{\bf Exceptional cases}

\medskip

{\bf 8. $AB(1,3)$} (also known as $F(4)$)

Here $V=V_{1}\oplus V_{2}$, where $V_1$ is a three dimensional
space with the basis $\{ e_{1},e_{2},e_{3}\}$
 and $V_2$ is a one-dimensional space generated by $e_4$.
The bilinear form $B$ is $$ B(u,v) = u^1v^1 + u^2v^2 + u^3v^3 -
3u^4v^4. $$

The root system $R$ is the set $$\pm e_{i}\pm e_{j},\quad i\ne j,
\quad \pm e_{i}, \quad i,j =1,2,3, \quad \pm e_4, \quad
\frac{1}{2}(\pm e_{1}\pm e_{2}\pm e_{3}\pm e_{4}),$$
$R_{re}=B_{3}\oplus A_{1}$, $R_{im}= \{\frac{1}{2}(\pm e_{1}\pm
e_{2}\pm e_{3}\pm e_{4}) \}$

\medskip
{\bf 9. $G(1,2)$} (also known as $G(3)$)

Here $V=V_{1}\oplus V_{2}$, where $V_1$ is a two-dimensional space,
generated by three vectors $e_1, e_2, e_3$ with the condition that
$e_{1}+e_{2}+e_{3}=0$ and $V_2$ is a one-dimensional space generated by $e_4$.
The form $B$ is determined by the following conditions:
$$
B(e_{i},e_{j})=-1 \quad \mbox{if}\quad i\ne j,\quad
B(e_{i},e_{i})=2,\quad
B(e_{i},e_{4})=0,  \quad
B(e_{4},e_{4})=-2,
$$
where $i,j = 1,2,3.$

$R$ consists of the vectors $\pm e_{i},(e_{i}-e_{j}),\pm e_{4},\pm
2e_{4},\pm e_{i}\pm e_{4}\quad, i\ne j, i,j \leq 3,$ $R_{re}=
G_{2}\oplus BC_{1}$, $R_{im}= \{\pm e_{i}\pm e_{4},i=1,2,3\}$
\medskip

{\bf 10. $D(2,1,\lambda)$}

Here $\lambda = (\lambda_1, \lambda_2, \lambda_3)$ are the parameters,
satisfying the relation
$\lambda_{1}+\lambda_{2}+\lambda_{3}=0$.

The space $V$ is $V_{1}\oplus V_{2}\oplus V_{3}$ the direct sum of
three one-dimensional spaces generated by $e_1, e_2, e_3$
respectively. The form $B$ is $$ B(u,v) = \lambda_1 u^1v^1 +
\lambda_2 u^2v^2 + \lambda_3 u^3v^3. $$

$R$ is the set $\{ \pm 2e_{1},\pm 2e_{2},\pm 2e_{3}, \ \pm
e_{1}\pm e_{2}\pm e_{3} \}$, $R_{re}=A_{1}\oplus A_{1}\oplus A_{1}$, $R_{im}= \{\pm
e_{1}\pm e_{2}\pm e_{3}\}$

Now we are going to explain how one can construct a family of the
Schr\"odinger operators related to a generalized root system $R
\subset V$.

These operators are defined on the same space $V$ and have the form
\begin{equation}
\label{dCMS}
L=-\Delta + \sum_{\alpha\in R_{+}}\frac{m_{\alpha}(m_{\alpha}+2m_{2\alpha}+1)(\alpha,\alpha)}{\sin^2(\alpha,x)}
\end{equation}
but now the brackets ( , ) and the Laplacian $\Delta$ correspond
to the new ("deformed") bilinear form $B$ on $V.$ Sometime we
would consider this operator on $V^*$ (which can be identified
with $V$ using $B$); in that case the brackets $(\alpha,x)$ should
be understood as a natural pairing between vectors and covectors.
The multiplicities $m_{\alpha}$ are related to $B$ in such a way
that the following conditions are satisfied:

1) the new form $B$ and the multiplicities are $W_{0}$-invariant;

2) all imaginary roots have the multiplicity 1;

3) the function $\psi_{0}=\prod_{\alpha\in R_{+}} \sin^{-m_{\alpha}}(\alpha,x)$ is a (formal) eigenfunction of
the corresponding Schr\"odinger operator (\ref{dCMS}).

We will call such forms $B$ and multiplicities {\it admissible}
and the corresponding operators (\ref{dCMS}) the {\it deformed CMS
operators related to generalized root system} $R$. If we replace
in (\ref{dCMS}) $x$ by $\omega x$ and let $\omega$ tend to $0$ (in
other words if we replace $\sin z$ by $z$) we will have the {\it
rational limits} of these operators.

Let us comment on the conditions 1)-3). The first one is very
natural: we would like to preserve the (partial) symmetry of the
system. The third condition is responsible for the existence of
the "radial gauge" of the operator $L$:
\begin{equation}
\label{radCMS} {\mathcal L}_{2}= - \Delta + 2\sum_{\alpha \in
R^{+}}m_{\alpha}\cot(\alpha,x)\partial_{\alpha}
\end{equation}
(see \cite{OP}, \cite{V}), and thus is motivated by the theory of
symmetric spaces. The second condition (related to condition 3) in
Serganova's axiomatics of the generalized root systems) looks very
simple but actually is the most difficult to justify. The
motivation comes from the theory of the {\it locus
configurations}, where the first examples of the such deformations
have been found \cite{CFV1},\cite{CFV2}. In that theory all the
multiplicities are integers and 1 is the smallest possible option.

A straightforward check shows that the condition 3) is equivalent to the following {\it main identity}:
\begin{equation}
\label{Main}
\sum\limits_{\alpha \not\sim \beta, \alpha,\beta\in R_{+}} m_{\alpha}m_{\beta}(\alpha, \beta) (\cot(\alpha,x) \cot(\beta,x)+1) \equiv 0,
\end{equation}
where $\alpha \not\sim \beta$ means that $\alpha$ is not
proportional to $\beta$ (note that in $BC(n,m)$ there are
proportional roots). In that case $L \psi_0 = \lambda \psi_0$ with
the eigenvalue $\lambda = |\rho(m)|^2$ where $\rho(m) =
\sum\limits_{\alpha \in R_{+}} m_{\alpha} \alpha$ (cf. \cite{OP},
\cite{V}).

To describe all possible deformations for a given generalized root
system one can use the fact that the condition (\ref{Main}) can be
checked separately for all two-dimensional subsystems (cf.
\cite{CFV2,V}). It is enough  to consider only the following
classical root systems $ A(n,m), BC(n,m)$ (from deformation point
of view others are simply special cases) and the exceptional root
systems $G(1,2), AB(1,3), D(2,1,\lambda )$. The forms $B$ are
obviously defined up to a multiple so we will choose some
normalization to avoid unnecessary constants.

A straightforward analysis leads to the following

\medskip
{\bf List of admissible deformations of generalized root systems.}
\medskip

Here $B$ is considered as an element of $V^* \otimes V^*$ which is
non-degenerate and thus determines an isomorphism between $V$ and
its dual $V^*.$ The formulas for the operators will be written on
$V^*$: this is more convenient for several reasons.

In all the cases the admissible forms depend on one parameter. We
denote this parameter $k$ and choose it in such a way that the
value $k=-1$ corresponds to the Lie superalgebra case.

\medskip
{\bf Classical series}
\medskip

{\bf $A(n,m)$}
\medskip

Form $B$ is
\begin{equation}
\label{kB} B = \sum_{i \in I_1} e^i\otimes e^i + k \sum_{j \in
I_2} e^j\otimes e^j,
\end{equation}
where $k$ is an arbitrary non-zero parameter. Multiplicities
$m_{\alpha} = m(\alpha)$ of the real roots are $$m(e_i - e_j) = k,
i,j \in I_1, \quad m(e_i - e_j) = k^{-1} , i,j \in I_2$$ (recall
that all imaginary roots have multiplicity 1).

Corresponding one-parameter family of the deformed CMS operators
has the form (\ref{anm}). We should mention that in the case $m=n$
the vector $v=\sum_{i\in I_{1}} e_{i} - \sum_{i\in I_{2}} e_{i}$
is not isotropic for the deformed form, so strictly speaking we
deform not generalized system of type $A(n-1,n-1)$ but its
(degenerate) extension.
\medskip

{\bf $BC(n,m)$}
\medskip

$B$ is the same as above, multiplicities are $$m(e_i \pm e_j) = k,
\quad m(e_i) = p, \quad m(2e_i) = q, i,j \in I_1,$$ $$ m(e_i \pm
e_j) = k^{-1}, \quad m(e_j) = r, \quad m(2e_j) = s, i,j \in I_2,$$
where $p,q,r,s$ are satisfying the relations $$p=kr,\quad
2q+1=k(2s+1).$$

Corresponding deformed CMS operators depend on three free
parameters and are given by (\ref{bcnm}).
\medskip

{\bf Exceptional cases}

\medskip

{\bf $AB(1,3)$}
\medskip

$$ B = e^1 \otimes e^1 + e^2 \otimes e^2 + e^3 \otimes e^3 + 3 k
e^4 \otimes e^4, $$ multiplicities are $$m(e_i) = a =
\frac{3k+1}{2} ,\quad  m(e_4) = b = \frac{1-k}{2k},\quad m(e_i \pm
e_j) = c = \frac{3k-1}{4}, i,j = 1,2,3.$$

Deformed CMS operator in this case has the form
\medskip
\begin{eqnarray}
\label{ab} L_{AB(1,3)}=-\left(\frac{\partial^2}{{\partial
x_{1}}^2}+\frac{\partial^2}{{\partial x_{2}}^2}
+\frac{\partial^2}{{\partial x_{3}}^2}\right)-3 k
\frac{\partial^2}{{\partial y}^2}+\sum_{i=1}^3
\frac{a(a+1)}{\sin^2x_{i}}+ \frac{3 k b(b+1)}{\sin^2y} \nonumber
\\ +\sum_{1\le i<j\le
3}\left(\frac{4c(c+1)}{\sin^2(x_{i}-x_{j})}+\frac{4c(c+1)}{\sin^2(x_{i}+x_{j})}\right)+
\frac{1}{4}\sum_{\pm}\frac{(3k+3)}{\sin^2\frac{1}{2}(y\pm x_{1}\pm
x_{2}\pm x_{3})}
\end{eqnarray}
where the parameters $a,b,c$ are given in terms of deformation
parameter $k$ above and the last sum is over all 8 possible
combinations of the signs. In this case we have only one free
parameter $k.$

\medskip

{\bf $G(1,2)$}
\medskip

In the basis $e_1, e_2, e_4$ the form $B$ has the form $$ B = e^1
\otimes e^1 + e^2 \otimes e^2 - \frac{1}{2}(e^1 \otimes e^2 + e^2
\otimes e^1) + k e^4 \otimes e^4.$$ Multiplicities are $$m(e_i)= a
= 1+2k,\quad m(e_i - e_j) = b = \frac{2k -1}{3}, \quad m(e_4) = c
= \frac{1}{k}+2,\quad m(2e_4) = d = \frac{1}{2k}-\frac{1}{2},$$
where $i,j = 1,2,3.$

Corresponding deformation of CMS operator has the form
\begin{eqnarray}
\label{g} L_{G(1,2)}&=& -\left(\frac{\partial^2}{{\partial
x_{1}}^2}-\frac{\partial^2}{\partial x_{1}\partial x_{2}}
+\frac{\partial^2}{{\partial x_{2}}^2}\right)- k
\frac{\partial^2}{{\partial y}^2}+\sum_{i=1}^3
\frac{a(a+1)}{\sin^2x_{i}}+\sum_{1\le i<j\le
3}\frac{3b(b+1)}{\sin^2(x_{i}-x_{j})} \nonumber \\& & +\frac{k
c(c+2d+1)}{\sin^2y}+\frac{4k d(d+1)}{\sin^22y}+\sum_{i=1}^3
\left(\frac{2(k+1)}{\sin^2(x_{i}-y)}+\frac{2(k+1)}{\sin^2(x_{i}+y)}\right)
\end{eqnarray}
where the parameters $a,b,c,d$ are given in terms of $k$ above.
Again we have one-parameter family.
\medskip

{\bf $D(2,1, \lambda)$}
\medskip

The form $B$ is $$ B = \lambda_1 e^1 \otimes e^1 + \lambda_2 e^2
\otimes e^2 + \lambda_3 e^3 \otimes e^3, $$ where $\lambda_i,\quad
i=1,2,3$ are arbitrary non-zero parameters. Let us introduce the
parameter $$k = \lambda_1+ \lambda_2 +\lambda_3 -1$$ so that when
$k = -1$ we have the Lie superalgebra case. The multiplicities
have the form $$ m(2 e_i) = m_{i}=\frac{k+1} {2\lambda_{i}}-1,
\quad i=1,2,3. $$

Corresponding deformed CMS operators are
\begin{equation}
\label{d}
L_{D(2,1,\lambda)}=\lambda_{1}\frac{\partial^2}{{\partial
x_{1}}^2}+\lambda_{2}\frac{\partial^2}{\partial x_{2}^2}
+\lambda_{3}\frac{\partial^2}{{\partial x_{3}}^2}+\sum_{i=1}^3
\frac{4\lambda_{i}m_{i}(m_{i}+1)}{\sin^22x_{i}}
+\sum_{\pm}\frac{2(k+1)}{\sin^2(x_{1}\pm x_{2}\pm x_{3})},
\end{equation}
where the last sum is again over all possible combinations of
signs (4 in this case). This family is living on the projective
plane with projective coordinates $\lambda_1: \lambda_2:
\lambda_3.$ The Lie superalgebra case corresponds to the line
$\lambda_1+ \lambda_2 +\lambda_3 = 0,$ so we have again only one
deformation parameter (say $k$).
\medskip

It is interesting to note that the potentials $U$ of all operators
listed above satisfy the so-called {\it locus conditions} (see
\cite{CFV2}): the first coefficient in the Laurent expansion of
$U$ in the direction of $\alpha$ is identically zero on the
hyperplanes $\sin(\alpha, x) = 0.$ For the real roots this is
obvious by symmetry reasons, but for imaginary roots this is not
and follows only from a direct case by case check. If we replace
the third condition for admissible deformations by these locus
relations we will have essentially the same list of the potentials
(some additional possibilities are due simply to the symmetry of
the potential under the change $m_{\alpha} \rightarrow
(-1-m_{\alpha})$). We do not have satisfactory explanation of this
phenomenon, which seems to be important (cf. \cite{V},
\cite{ChV}).

Since the locus configurations (in the rational case) are known to
be related to the Huygens' Principle (see \cite{CFV2}) it is
natural to ask about the deformed CMS operators with all
multiplicities being integers. A simple check of the cases listed
above shows that the only non-trivial cases correspond to the
systems $A(n,1)$ and $BC(n,1)$ which has already been discovered
in \cite{CFV1},\cite{CFV2}. This indicates that the list of known
locus configurations \cite{ChV} could be in fact complete and
therefore could give the answer to the famous Hadamard problem in
the theory of Huygens' principle.

\section{Construction of quantum integrals for classical series.}

In this section we present a recursive formula for the quantum
integrals of the deformed CMS problems (\ref{anm}),(\ref{bcnm})
(more precisely, for its hyperbolic versions). Our formula is a
properly deformed version of the formula used by A. Matsuo in his
paper \cite{Matsuo} on the relations between CMS quantum problem
and KZ equation. We would like to mention also that for $m=1$ an
equivalent set of integrals was found earlier in \cite{CFV1} using
very different ideas.

Let $V$ be $n+m$-dimensional vector space with the deformed inner
product $$ (u,v) = \sum_{i=1}^n u^i v^i + k \sum_{j=n+1}^{n+m} u^j
v^j$$ corresponding to the form $B$ given by (\ref{kB}).

When $k=1$ we have the standard Euclidean scalar product which we
denote by $<,>:$ $$ <u,v> = u^1 v^1 + ... +u^{n+m} v^{n+m}.$$ We
will be using the fact that the generalized root systems $R$ of
type $A(n-1,m-1)$ and $BC(n,m)$ with respect to the scalar product
$<u,v>$ coincide with the usual root systems $A(n+m-1)$ and
$BC(n+m)$ (see previous section). The corresponding Weyl groups
$W$ are generated by reflections with respect to the Euclidean
form which we denote as $s_{<\alpha>}.$

Let us introduce the operator ${\mathcal B}$ by the relation $$
B(u,v) = <{\mathcal B}u, v>: $$ ${\mathcal B} e_i = e_i,
i=1,...,n, \quad {\mathcal B} e_j = k e_j, j = n+1, ..., n+m.$ We
will call a vector $v \in V$ {\it homogeneous} if it is an
eigenvector of ${\mathcal B}$ (and thus for any pair of forms in
our family).

For any homogeneous vector $v$ the following relation holds: $$
\frac{(x,v)}{<x,v>}=\frac{(v,v)}{<v,v>} $$ for any vector $x \in
V.$ Obviously in our case the set of homogeneous vectors is $V_1
\cup V_2,$ where $V_1$ and $V_2$ are generated by the first $n$
and last $m$ basic vectors respectively. It will be important for
us that for the classical systems there exists an orbit ${\mathcal
O}$ of the Weyl group $W$, which consists of homogeneous vectors.
Indeed, for $A(n-1,m-1)$ root system one can take ${\mathcal O} =
{e_i}$ and for $BC(n,m)$ such an orbit is ${\mathcal O} = {\pm
e_i}.$

In this section we will assume that $x \in V$ and the brackets
$(\alpha,x)$ denote the deformed product given by $B$.

Let us define for  $\alpha\in R$ the following functions on $V$ $$
f_{\alpha}(x)=\frac{1}{2}\frac{e^{(\alpha,x)}+1}{e^{(\alpha,x)}-1}
= \frac{1}{2} \coth \frac{(\alpha,x)}{2},\quad \varphi_{\alpha}(x)
= \frac{1}{4}-f_{\alpha}(x)^2 = - \frac{1}{4 \sinh^2
\frac{(\alpha,x)}{2}}. $$ They satisfy the following relations: $$
{\partial}_{v}f_{\alpha} = (v,\alpha) \varphi_{\alpha}, \quad
{\partial}_{v} \varphi_{\alpha} = - 2(v,\alpha) f_{\alpha}
\varphi_{\alpha} $$ for any $v \in V.$

Let us define now for any natural number $p$ the operator ${\partial}_{v}^{(p)}$
by the following recurrent procedure
$$
{\partial}_{v}^{(1)}={{\partial}_{v}},
$$
\begin{equation}
\label{recurrence}
{\partial}_{v}^{(p)}={{\partial}_{v}}{\partial}_{v}^{(p-1)}-
\sum_{\alpha \in R^{+}} m_{\alpha}(\alpha,v)f_{\alpha} (
  {\partial}_{v}^{(p-1)}-{\partial}_{s_{<\alpha>} v}^{(p-1)}),
\end{equation}
where $R^+$ is a positive part of $R$ and $s_{<\alpha>}$ is the reflection corresponding to the root
$\alpha$ with respect to the Euclidean form $< , >$. The formula (\ref{recurrence})
is a deformed version of the formula (2.3.6) from A. Matsuo's paper \cite{Matsuo}.

Take now an orbit $\mathcal O$ of the corresponding Weyl group $W$
consisting of homogeneous elements (see above) and define
\begin{equation}
\label{integrals} {\mathcal L}_{p} =\sum_{v\in \mathcal
O}\frac{{\partial}_{v}^{(p)}}{(v,v)}.
\end{equation}

One can easily check using the relation $$
{\partial}_{v}-{\partial}_{s_{<\alpha>}
v}=2\frac{<\alpha,v>}{<\alpha,\alpha>}\partial_{\alpha} $$ that $$
{\mathcal L}_{2}=\sum_{v\in \mathcal
O}\frac{{\partial}_{v}^{2}}{(v,v)}- 2\sum_{\alpha \in R^{+}}
m_{\alpha}f_{\alpha}{\partial}_{\alpha} $$ is up to a coefficient
the deformed CMS operator (\ref{radCMS}) (in the hyperbolic
version and radial gauge) after a scaling $x \rightarrow 2x$.

\medskip

{\bf Theorem 1.} {\it The operators ${\mathcal L}_{p}$ given by
the formula (\ref{integrals}) commute with each other}:
$$[{\mathcal L}_{p},{\mathcal L}_{q}] = 0$$ {\it and thus are the
quantum integrals of the corresponding deformed CMS problem
(\ref{radCMS}) related to classical generalized root systems.}

\medskip

The proof is based on the following

\medskip

{\bf Proposition 1.} {\it The operators ${\partial}_{v}^{(p)}$
satisfy the following commutation relation with the deformed CMS
operator:}
\begin{equation}
\label{0.2} \left [ {\mathcal L}_{2},{\partial}_{v}^{(p)}\right
]=(v,v) \sum_{\alpha \in R^{+}}
m_{\alpha}\frac{<\alpha,\alpha>}{<v,v>} \varphi_{\alpha}
({\partial}_{v}^{(p)}-{\partial}_{s_{<\alpha>} v}^{(p)})
\end{equation}

\medskip

\noindent The proof is by induction in $p$. For $p=1$ we have the
relation
\begin{equation}
\label{rf} \left [ {\mathcal L}_{2},{\partial}_{v}\right ]=(v,v)
\sum_{\alpha \in R^{+}} m_{\alpha}\frac{<\alpha,\alpha>}{<v,v>}
\varphi_{\alpha} ({\partial}_{v}-{\partial}_{s_{<\alpha>} v}),
\end{equation}
which is easy to check.

The proof of the induction step is a long but straightforward
calculation. We reproduce the main steps to show the role of the
properties of admissible deformations here.

Let us assume that the statement is true for all natural numbers
less than $p$. We have $$ \left [ {\mathcal
L}_{2},{\partial}_{v}^{(p)}\right ]= \left [ {\mathcal
L}_{2},{\partial}_{v}\right ]{\partial}_{v}^{(p-1)}+
{\partial}_{v}\left [ {\mathcal
L}_{2},{\partial}_{v}^{(p-1)}\right ]- \sum_{\alpha \in R^{+}}
m_{\alpha}(\alpha,v)\left [{\mathcal L}_{2},f_{\alpha}\right]
  ({\partial}_{v}^{(p-1)}-{\partial}_{s_{<\alpha>} v}^{(p-1)})
$$
$$ -\sum_{\alpha \in R^{+}} m_{\alpha}(\alpha,v)f_{\alpha}
  \left [{\mathcal L}_{2},{\partial}_{v}^{(p-1)}-{\partial}_{s_{<\alpha>} v}^{(p-1)}
\right ]=(v,v) \sum_{\alpha \in R^{+}}
m_{\alpha}\frac{<\alpha,\alpha>}{<v,v>} \varphi_{\alpha}
({\partial}_{v}-{\partial}_{s_{<\alpha>} v})
{\partial}_{v}^{(p-1)}
$$ $$ +(v,v)\partial_{v}\sum_{\alpha \in
R^{+}} m_{\alpha}\frac{<\alpha,\alpha>}{<v,v>} \varphi_{\alpha}
({\partial}_{v}^{(p-1)}-{\partial}_{s_{<\alpha>} v}^{(p-1)})-
2\sum_{\alpha \in R^{+}} m_{\alpha}(\alpha,v)\varphi_{\alpha}
\partial_{\alpha}({\partial}_{v}^{(p-1)}-{\partial}_{s_{<\alpha>} v}^{(p-1)})
$$ $$ +2\sum_{\alpha \in R^{+}} m_{\alpha}(\alpha,v)f_{\alpha}
\partial_{\alpha}(f_{\alpha})
({\partial}_{v}^{(p-1)}-{\partial}_{s_{<\alpha>} v}^{(p-1)})-
\sum_{\alpha \in R^{+}} m_{\alpha}(\alpha,v)f_{\alpha}
  \left [{\mathcal L}_{2},{\partial}_{v}^{(p-1)}-{\partial}_{s_{<\alpha>} v}^{(p-1)}
\right ] $$ $$ +2\sum_{\alpha,\beta \in R^{+}} m_{\alpha}m_{\beta}
(\alpha,v)f_{\beta}\partial_{\beta}(f_{\alpha})
({\partial}_{v}^{(p-1)}-{\partial}_{s_{<\alpha>} v}^{(p-1)}), $$
where we have used the induction assumption and the relation $$
\left [ {\mathcal L}_{2},f_{\alpha}\right
]=2\varphi_{\alpha}\partial_{\alpha}
-2f_{\alpha}\partial_{\alpha}(f_{\alpha})-2\sum_{\beta \in R^+}
m_{\beta}f_{\beta}\partial_{\beta}(f_{\alpha}). $$ Let us denote
the sum of the last two sums in the previous expression as $B$: $$
B = 2\sum_{\alpha,\beta \in R^{+}} m_{\alpha}m_{\beta}
(\alpha,v)f_{\beta}\partial_{\beta}(f_{\alpha})
({\partial}_{v}^{(p-1)}-{\partial}_{s_{<\alpha>} v}^{(p-1)})-
\sum_{\alpha \in R^{+}} m_{\alpha}(\alpha,v)f_{\alpha}
  \left [{\mathcal L}_{2},{\partial}_{v}^{(p-1)}-{\partial}_{s_{<\alpha>} v}^{(p-1)}
\right ] $$ and the rest of the previous expression as $A.$ Using
the homogeneity of $v$ we can rewrite $A$ as

$$
A = (v,v)\sum_{\alpha \in R^{+}} m_{\alpha}\frac{<\alpha,\alpha>}{<v,v>}
\varphi_{\alpha}\left \{({\partial}_{v}-{\partial}_{s_{<\alpha>} v})
\partial_{v}^{(p-1)}+\partial_{v}
  ({\partial}_{v}^{(p-1)}-{\partial}_{s_{<\alpha>} v}^{(p-1)})\right \}
$$ $$ +(v,v)\sum_{\alpha \in R^{+}}
m_{\alpha}\frac{<\alpha,\alpha>}{<v,v>}
\partial_{v}(\varphi_{\alpha}) ({\partial}_{v}^{(p-1)}-
{\partial}_{s_{<\alpha>} v}^{(p-1)})-
2\sum_{\alpha \in R^{+}} m_{\alpha}(\alpha,v)\varphi_{\alpha}
\partial_{\alpha}({\partial}_{v}^{(p-1)}-{\partial}_{s_{<\alpha>} v}^{(p-1)})
$$ $$ +2\sum_{\alpha \in R^{+}} m_{\alpha}(\alpha,v)f_{\alpha}
\partial_{\alpha}(f_{\alpha})
({\partial}_{v}^{(p-1)}-{\partial}_{s_{<\alpha>} v}^{(p-1)})
$$
$$
=(v,v)\sum_{\alpha \in R^{+}} m_{\alpha}\frac{<\alpha,\alpha>}{<v,v>}
\varphi_{\alpha}({\partial}_{v}\partial_{v}^{(p-1)}-
  {\partial}_{s_{\alpha }v}{\partial}_{s_{<\alpha>} v}^{(p-1)})
$$ $$ +2\sum_{\alpha \in R^{+}} m_{\alpha}(\alpha,v)f_{\alpha}
(\partial_{\alpha}(f_{\alpha})-\frac{(v,v)}{<v,v>}<\alpha,\alpha>
\varphi_{\alpha})({\partial}_{v}^{(p-1)}-{\partial}_{s_{<\alpha>}
v}^{(p-1)}). $$
Now let us use that
$$
{{\partial}_{v}}{\partial}_{v}^{(p-1)}={\partial}_{v}^{(p)}+
\sum_{\beta \in R^{+}} m_{\beta}(\beta,v)f_{\beta} (
  {\partial}_{v}^{(p-1)}-{\partial}_{s_{<\beta>} v}^{(p-1)}),
$$
$$
{{\partial}_{s_{<\alpha>}v}}{\partial}_{s_{<\alpha>}v}^{(p-1)}=
{\partial}_{s_{<\alpha>}v}^{(p)}+\sum_{\beta \in R^{+}} m_{\beta}
(\beta,s_{<\alpha>}v)f_{\beta}
  ({\partial}_{s_{<\alpha>}v}^{(p-1)}-{\partial}_{s_{<\beta>}s_{<\alpha>}v}^{(p-1)}),
$$ to rewrite the last expression as $$ A = (v,v)\sum_{\alpha \in
R^{+}} m_{\alpha}\frac{<\alpha,\alpha>}{<v,v>}
\varphi_{\alpha}(\partial_{v}^{(p)}-{\partial}_{s_{<\alpha>}
v}^{(p)}) $$ $$ +(v,v)\sum_{\alpha \in R^{+}}
m_{\alpha}\frac{<\alpha,\alpha>}{<v,v>}
\varphi_{\alpha}\sum_{\beta \in R^{+}} m_{\beta}(\beta,v)f_{\beta}
(
  {\partial}_{v}^{(p-1)}-{\partial}_{s_{<\beta>} v}^{(p-1)})
$$ $$ -(v,v)\sum_{\alpha \in R^{+}}
m_{\alpha}\frac{<\alpha,\alpha>}{<v,v>}
\varphi_{\alpha}\sum_{\beta \in R^{+}} m_{\beta}
(\beta,s_{<\alpha>}v)f_{\beta}
  ({\partial}_{s_{<\alpha>}v}^{(p-1)}-{\partial}_{s_{<\beta>}s_{<\alpha>}v}^{(p-1)})
$$ $$ +2\sum_{\alpha \in R^{+}} m_{\alpha}(\alpha,v)f_{\alpha}
(\partial_{\alpha}(f_{\alpha})-\frac{(v,v)}{<v,v>}<\alpha,\alpha>
\varphi_{\alpha})({\partial}_{v}^{(p-1)}-{\partial}_{s_{<\alpha>}
v}^{(p-1)}) $$ $$=(v,v)\sum_{\alpha \in R^{+}}
m_{\alpha}\frac{<\alpha,\alpha>}{<v,v>}
\varphi_{\alpha}(\partial_{v}^{(p)}-{\partial}_{s_{<\alpha>}
v}^{(p)}) $$ $$+ (v,v)\sum_{\alpha \in R^{+}}
m_{\alpha}\frac{<\alpha,\alpha>}{<v,v>}
\varphi_{\alpha}m_{\alpha}f_{\alpha}\left
((\alpha,v)+(\alpha,s_{<\alpha>}v) \right )
(\partial_{v}^{(p-1)}-{\partial}_{s_{<\alpha>} v}^{(p-1)}) $$ $$
+2\sum_{\alpha \in R^{+}}
m_{\alpha}(\alpha,v)f_{\alpha}\varphi_{\alpha} \left
((\alpha,\alpha)-\frac{(v,v)}{<v,v>}<\alpha,\alpha>\right )
({\partial}_{v}^{(p-1)}-{\partial}_{s_{<\alpha>} v}^{(p-1)})$$ $$
+(v,v)\sum_{\alpha \in R^{+}}
m_{\alpha}\frac{<\alpha,\alpha>}{<v,v>}
\varphi_{\alpha}\sum_{\beta \in R^{+}, \beta\ne\alpha}
m_{\beta}(\beta,v)f_{\beta} (
  {\partial}_{v}^{(p-1)}-{\partial}_{s_{<\beta>} v}^{(p-1)})
$$ $$ -(v,v)\sum_{\alpha \in R^{+}}
m_{\alpha}\frac{<\alpha,\alpha>}{<v,v>}
\varphi_{\alpha}\sum_{\beta \in R^{+}, \beta\ne\alpha} m_{\beta}
(\beta,s_{<\alpha>}v)f_{\beta}
({\partial}_{s_{<\alpha>}v}^{(p-1)}-{\partial}_{s_{<\beta>}s_{<\alpha>}v}^{(p-1)}).
$$ Combining the second and third sums we come to the following
expression $$ \left
(m_{\alpha}\{(\alpha,v)+(\alpha,s_{<\alpha>}v)\}+
2<\alpha,v>\{\frac{(\alpha,\alpha)}{<\alpha,\alpha>}-\frac{(v,v)}{<v,v>}\}
\right ), $$ which can be rewritten in the form $$ -2<\alpha, v>
(m_{\alpha}-1) \left (\frac{(\alpha,\alpha)} {<\alpha,\alpha>}
-\frac{(v,v)}{<v,v>}\right  ). $$ We claim that this is $0$ for
any root $\alpha.$ Indeed if $\alpha$ is imaginary then
$m_{\alpha} =1$ by our assumption (property 2 of admissible
deformations). If $\alpha$ is real and $<\alpha,v>$ is not zero,
then $\left (\frac{(\alpha,\alpha)}
{<\alpha,\alpha>}-\frac{(v,v)}{<v,v>}\right  ) = 0$ for any $v$
from our orbit.

Thus we come to the following expression for $A:$ $$ A =
(v,v)\sum_{\alpha \in R^{+}}
m_{\alpha}\frac{<\alpha,\alpha>}{<v,v>}
\varphi_{\alpha}(\partial_{v}^{(p)}-{\partial}_{s_{<\alpha>}
v}^{(p)}) $$ $$ +(v,v)\sum_{\alpha \in R^{+}}
m_{\alpha}\frac{<\alpha,\alpha>}{<v,v>}
\varphi_{\alpha}\sum_{\beta \in R^{+}\beta\ne\alpha}
m_{\beta}(\beta,v)f_{\beta}
  {\partial}_{v}^{(p-1)}-{\partial}_{s_{<\beta>} v}^{(p-1)})
$$ $$ -(v,v)\sum_{\alpha \in R^{+}}
m_{\alpha}\frac{<\alpha,\alpha>}{<v,v>}
\varphi_{\alpha}\sum_{\beta \in R^{+}\beta\ne\alpha} m_{\beta}
(\beta,s_{<\alpha>}v)f_{\beta}
  ({\partial}_{s_{<\alpha>}v}^{(p-1)}-{\partial}_{s_{<\beta>}s_{<\alpha>}v}^{(p-1)}).
$$
Now let us look at term $B.$ We have
$$
\sum_{\alpha \in R^{+}} m_{\alpha}(\alpha,v)f_{\alpha}
  \left [{\mathcal L}_{2},{\partial}_{v}^{(p-1)}-{\partial}_{s_{<\alpha>} v}^{(p-1)}
\right ] $$ $$ =\sum_{\alpha ,\beta \in R^{+}}
m_{\alpha}(\alpha,v)f_{\alpha}m_{\beta}<\beta,\beta>
\varphi_{\beta}
  \frac{(v,v)}{<v,v>}({\partial}_{v}^{(p-1)}-{\partial}_{s_{<\beta>} v}
^{(p-1)}) $$ $$-\sum_{\alpha ,\beta \in R^{+}}
m_{\alpha}(\alpha,v)f_{\alpha}m_{\beta}<\beta,\beta>
\varphi_{\beta}
\frac{(s_{<\alpha>}v,s_{<\alpha>}v)}{<s_{<\alpha>}v,s_{<\alpha>}v>}
({\partial}_{s_{<\alpha>}v}^{(p-1)}-{\partial}_{s_{<\beta>}
s_{<\alpha>}v}^{(p-1)}) $$
$$ =\sum_{\alpha ,\beta \in
R^{+}\beta\ne\alpha} m_{\alpha}(\alpha,v)
f_{\alpha}m_{\beta}<\beta,\beta>\varphi_{\beta}
  \frac{(v,v)}{<v,v>}({\partial}_{v}^{(p-1)}-{\partial}_{s_{<\beta>} v}
^{(p-1)}) $$
$$-\sum_{\alpha ,\beta \in R^{+}\beta\ne\alpha}
m_{\alpha}(\alpha,v)
f_{\alpha}m_{\beta}<\beta,\beta>\varphi_{\beta}
\frac{(s_{<\alpha>}v,s_{<\alpha>}v)}{<s_{<\alpha>}v,s_{<\alpha>}v>}
({\partial}_{s_{<\alpha>}v}^{(p-1)}-{\partial}_{s_{<\beta>}
s_{<\alpha>}v}^{(p-1)})$$
$$+ \sum_{\alpha \in R^{+}}
m_{\alpha}^2(\alpha,v) f_{\alpha}<\alpha,\alpha>\varphi_{\alpha}
  \left (\frac{(v,v)}{<v,v>}+\frac{(s_{<\alpha>}v,s_{<\alpha>}v)}
{<s_{<\alpha>}v,s_{<\alpha>}v>}\right )
({\partial}_{v}^{(p-1)}-{\partial}_{s_{<\alpha>} v}^{(p-1)}).
$$
Combining the last term with the term in the second sum of $B$
corresponding to $\beta = \alpha$ and using the relation
$$
(s_{<\alpha>}v,s_{<\alpha>}v)=(v,v)+4\frac{<\alpha,v>^2}{<\alpha,\alpha>}
\left (\frac{(\alpha,\alpha)}{<\alpha,\alpha>}-\frac{(v,v)}{<v,v>}\right )
$$
we have
$$
m_{\alpha}^{2}(\alpha,v)f_{\alpha}\left\{2\partial_{\alpha}(f_{\alpha})
-<\alpha,\alpha>\varphi_{\alpha}
  \left (\frac{(v,v)}{<v,v>}+\frac{(s_{<\alpha>}v,s_{<\alpha>}v)}
{<s_{<\alpha>}v,s_{<\alpha>}v>}\right )\right\}
({\partial}_{v}^{(p-1)}-{\partial}_{s_{<\alpha>} v}^{(p-1)}),
$$
which is zero since
$$
2\left (1-2\frac{<\alpha,v>^2}{<\alpha,\alpha><v,v>}\right )
\left (\frac{(\alpha,\alpha)}{<\alpha,\alpha>}-
\frac{(v,v)}{<v,v>}\right )({\partial}_{v}^{(p-1)}-{\partial}_{s_{<\alpha>} v}^{(p-1)})=0
$$
for any root: for the real roots the product of the last two brackets is zero while
for imaginary roots the first bracket is zero.

Thus we arrive at the following expression for $$ \left [
{\mathcal L}_{2},{\partial}_{v}^{(p)}\right ]= A + B =
(v,v)\sum_{\alpha \in R^{+}}
m_{\alpha}\frac{<\alpha,\alpha>}{<v,v>}
\varphi_{\alpha}(\partial_{v}^{(p)}-{\partial}_{s_{<\alpha>}
v}^{(p)}) $$ $$ +(v,v)\sum_{\beta\ne\alpha} m_{\alpha}
m_{\beta}\frac{<\alpha,\alpha>}
{<v,v>}\varphi_{\alpha}f_{\beta}\left\{(\beta,v)
  ({\partial}_{v}^{(p-1)}-{\partial}_{s_{<\beta>} v}^{(p-1)})-(\beta,s_{<\alpha>}v)
  ({\partial}_{s_{<\alpha>}v}^{(p-1)}-{\partial}_{s_{<\beta>}s_{<\alpha>}v}^{(p-1)})
\right\}-
$$
$$
\sum_{\beta\ne\alpha} m_{\alpha}m_{\beta}(\alpha,v)\frac{<\beta,\beta>}{<v,v>}f_{\alpha}\varphi_{\beta}
\left \{(v,v)({\partial}_{v}^{(p-1)}-{\partial}_{s_{<\beta>} v}^{(p-1)})-
(s_{<\alpha>}v,s_{<\alpha>}v)
({\partial}_{s_{<\alpha>}v}^{(p-1)}-{\partial}_{s_{<\beta>} s_{<\alpha>}v}^{(p-1)})
\right \}
$$
$$
+ 2\sum_{\alpha\ne\beta } m_{\alpha}m_{\beta}
(\alpha,v)f_{\beta}\partial_{\beta}(f_{\alpha})
({\partial}_{v}^{(p-1)}-{\partial}_{s_{<\alpha>} v}^{(p-1)}).
$$
Let us denote the sum of the last three sums in the previous expression as $C.$
We must show that $C$ is identically zero.
Let us notice that
$$
\sum_{\alpha\ne\beta } m_{\alpha}m_{\beta}
(\alpha,v)f_{\beta}\partial_{\beta}(f_{\alpha})=
\sum_{\alpha\ne\beta } m_{\alpha}m_{\beta}
(\alpha,v)f_{\beta}(\beta,\alpha)\varphi_{\alpha}=
\partial_{v}\left(\sum_{\alpha\ne\beta } m_{\alpha}m_{\beta}
(\alpha,\beta)f_{\beta}f_{\alpha}\right) $$ But according to our
assumption (third property of admissible deformations)
$$
\sum_{\alpha\ne\beta } m_{\alpha}m_{\beta}
(\alpha,\beta)f_{\beta}f_{\alpha}=const+4\sum_{\alpha }
  m_{\alpha}m_{2\alpha}
(\alpha,\alpha)f_{2\alpha}f_{\alpha} $$ (see the formula
(\ref{Main}) above). Thus $C$ can be rewritten as
$$\sum_{\beta\ne\alpha} (v,v) m_{\alpha}
m_{\beta}\frac{<\alpha,\alpha>}
{<v,v>}\varphi_{\alpha}f_{\beta}\left\{(\beta,v)
  ({\partial}_{v}^{(p-1)}-{\partial}_{s_{<\beta>} v}^{(p-1)})-(\beta,s_{<\alpha>}v)
  ({\partial}_{s_{<\alpha>}v}^{(p-1)}-{\partial}_{s_{<\beta>}s_{<\alpha>}v}^{(p-1)})
\right\}$$ $$-\sum_{\beta\ne\alpha} m_{\alpha}m_{\beta}(\alpha,v)
\frac{<\beta,\beta>}{<v,v>}f_{\alpha}\varphi_{\beta}\left \{
(v,v)({\partial}_{v}^{(p-1)}-{\partial}_{s_{<\beta>} v}^{(p-1)})-
(s_{<\alpha>}v,s_{<\alpha>}v)
({\partial}_{s_{<\alpha>}v}^{(p-1)}-{\partial}_{s_{<\beta>}
s_{<\alpha>}v}^{(p-1)}) \right \} $$ $$ -2\sum_{\alpha\ne\beta
}m_{\alpha}m_{\beta}(\alpha,v)f_{\beta}\partial_{\beta}(f_{\alpha})
{\partial}_{s_{<\alpha>} v}^{(p-1)} + 4\partial_{v}(\sum_{\alpha
}m_{\alpha}m_{2\alpha}(\alpha,\alpha)f_{2\alpha}f_{\alpha})\partial_{v}^{(p-1)}
$$ $$= \sum_{\alpha\ne\beta } S_1(\alpha, \beta, v) -
\sum_{\alpha\ne\beta } S_2(\alpha, \beta, v) -
\sum_{\alpha\ne\beta } S_3(\alpha, \beta, v) + \sum_{\alpha}
S_4(\alpha, v), $$ where $S_1, S_2, S_3, S_4$ denote the terms in
the first, second, third and forth sums respectively. Choose in
the first three sums the terms with $\beta = 2\alpha$. We have
$\sum_{\alpha} S_1(\alpha, 2\alpha, v) = \sum_{\alpha}
S_1(2\alpha, \alpha, v) = 0$ and $$ -\sum_{\alpha} (S_2(\alpha,
2\alpha, v) + S_2 (2\alpha, \alpha, v)) - \sum_{\alpha}
(S_3(\alpha, 2\alpha, v) + S_3 (2\alpha, \alpha, v)) +
\sum_{\alpha} S_4(\alpha, v) = 0. $$ Thus we must show only that
$$ C= \sum\limits_{\alpha \not\sim \beta, \alpha,\beta\in R_{+}}
(S_1(\alpha, \beta, v) - S_2(\alpha, \beta, v) - S_3(\alpha,
\beta, v)) =0. $$ This can be done separately for each
two-dimensional subsystems (cf. \cite{CFV2}). Proposition 1 is
proven.

Now we are ready to prove Theorem 1. When $p=2$ (i.e. when
${\mathcal L}_{p} = {\mathcal L}_{2} = {\mathcal L}$ is the
deformed CMS operator) this follows immediately from the Lemma.
Indeed $$[{\mathcal L}_{2}, {\mathcal L}_{p}] = \sum_{v\in
\mathcal O}[{\mathcal L}_{2},\frac{{\partial}_{v}^{(p)}}{(v,v)}] =
\sum_{v\in \mathcal O}\sum_{\alpha \in R^{+}}
m_{\alpha}\frac{<\alpha,\alpha>}{<v,v>} \varphi_{\alpha}
({\partial}_{v}^{(p)}-{\partial}_{s_{<\alpha>} v}^{(p)}),$$ which
obviously is identically zero.

To prove that these operators commute for any $p, q$ we borrow the
idea from T. Oshima's paper \cite{Oshima}. Consider an involution
$\sigma$ on the space of all differential operators on $V$
corresponding to the change $x \rightarrow -x$ and the standard
anti-involution $*$: operator $L^*$ is a formal adjoint to $L$. We
have $[L_1^{\sigma}, L_2^{\sigma}] = [L_1, L_2]^{\sigma}$ and
$[L_1^{*}, L_2^{*}] = -[L_1, L_2]^{*}.$ Our operators ${\mathcal
L}_{p}$ have the following properties with respect to these
involutions: ${\mathcal L}_{p}^* = {\mathcal L}_{p}^{\sigma} =
(-1)^p {\mathcal L}_{p}.$

Now let us consider the commutator $C = [{\mathcal
L}_{p},{\mathcal L}_{q}]$. By Jacobi identity $[C,{\mathcal
L}_{2}] = 0$, so we can use Berezin's lemma \cite{Berezin} which
says that in such a case the highest symbol of $C$ must be
polynomial in $x.$ Since in our case it must also be periodic this
implies that the highest symbol is constant. We claim that it is
actually zero. Indeed $C^* = [{\mathcal L}_{p},{\mathcal L}_{q}]^*
= -[{\mathcal L}_{p}^*,{\mathcal L}_{q}^*] = -[{\mathcal
L}_{p}^{\sigma},{\mathcal L}_{q}^{\sigma}] = - [{\mathcal
L}_{p},{\mathcal L}_{q}]^{\sigma} = - C^{\sigma},$ so $C^* = -
C^{\sigma}.$ Looking at the highest symbol in this relation we see
that it must be zero. This completes the proof of the Theorem 1.

Notice that as follows from the formula for the integrals in the
$BC_{n,m}$ case all the integrals ${\mathcal L}_p$ with odd $p$
are actually vanish, so in that case we will consider only even
$p$.

Recall now that the quantum system in ${\bf R}^n$ is called {\it
integrable} if it has at least $n$ commuting independent quantum
integrals.

{\bf Corollary.} {\it Deformed CMS problems (\ref{anm}),
(\ref{bcnm}) related to the classical generalized root systems are
integrable. The same is true for their rational limits.}

To have the integrals in the rational limit one should replace in
the formulas of this section $\sinh z$ by $z$ and $\coth z$ by
$z^{-1},$ so that $f_{\alpha}(x) = (\alpha,x)^{-1},
\varphi_{\alpha}(x) = -(\alpha,x)^{-2}.$

\section {Algebra $\Lambda_{R,B}$ and Harish-Chandra homomorphism.}

Let $R \subset V$ be a classical generalized root system and $(R,
m, B)$ be its admissible deformation described in the section 1.
Let us introduce the corresponding algebra
$\Lambda^{\omega}_{R,B}$ as the algebra of polynomial functions
$p(x)$ on $V,$ which satisfy the following properties (cf.
\cite{CV}):

1) $p(x)$ are invariant with respect to Weyl group $W_0,$
corresponding to the real roots of the system;

2) $p(x + \omega\alpha) \equiv p(x - \omega\alpha)$ on the
hyperplane $(\alpha,x) = 0$ for any imaginary root $\alpha,$ where
$(\alpha,x)$ is the deformed scalar product determined by $B$.

In the limit $\omega \to 0$ we have the algebra
$\Lambda^{0}_{R,B}$ of $W_0$-invariant polynomials with the
properties 1) and

$2)^0$ $\partial_{\alpha} p(x) \equiv 0$ on the hyperplane
$(\alpha,x) = 0$ for any imaginary root $\alpha \in R.$

One can consider this algebra also as a subalgebra of polynomial
functions on $V^*$ satisfying the same relation $2)^0$ where
$(\alpha,x)$ is understood as pairing between vector and covector
and $\partial_{\alpha}$ is defined using the deformed form $B.$
Below we will be using this realization.

Since the algebras $\Lambda^{\omega}_{R,B}$ are obviously
isomorphic for all $\omega \neq 0$ we will assume later on that
$\omega = 1/2$ considering only algebras $\Lambda_{R,B} =
\Lambda^{1/2}_{R,B}$ and $\Lambda^{0}_{R,B}$. We are going to show
that for generic $k$ these two algebras are actually isomorphic to
the algebras generated by the quantum integrals of the deformed
CMS problems from the previous section in trigonometric and
rational case respectively.

{\bf Remark.} We should mention that in the case when all the
multiplicities are integer the algebra of quantum integrals is
actually much bigger and is called algebra of {\it
quasi-invariants}, see \cite{CFV2}, \cite{FV}). For example when
$k=1$ the quasi-invariants are polynomials satisfying the property
$2)^0$ for {\it all} roots, but no symmetry is imposed (see
\cite{FV2} for the latest results in this direction).

It is obvious that the highest order component of any polynomial
$P \in \Lambda_{R,B}$ belongs to the algebra $\Lambda^0_{R,B}.$
More subtle question is whether for any homogeneous $Q \in
\Lambda^0_{R,B}$ there exists $P \in \Lambda_{R,B}$ such that $Q$
is the highest term of $P.$ We will show that at least for generic
values of the deformation parameter $k$ this is true, which means
that $\Lambda^{0}_{R,B}$ is the associated graded algebra for
$\Lambda_{R,B}$.

We are going now to describe the algebras $\Lambda^{0}_{R,B}$ more
explicitly. Let us start with the type $A(n-1,m-1)$. The
corresponding algebra can be realized as the following algebra
$\Lambda^0_{n,m;k} \subset {\bf C}[V^*] = {\bf
C}[x_{1},\dots,x_{n},y_{1},\dots,y_{m}]$ consisting of the
polynomials $f (x_{1},\dots,x_{n},y_{1},\dots,y_{m})$ which are
symmetric in $x_{1},\dots,x_{n}$ and $y_{1},\dots,y_{m}$
separately and satisfy the conditions $$
(\frac{\partial}{\partial{x_{i}}}-k\frac{\partial}{
\partial{y_{j}}} )f \equiv 0$$
on each hyperplane $x_{i}- y_{j} = 0$ for $i=1,...,n$ and
$j=1,...,m.$

It is very easy to check that the {\it deformed Newton sums}
\begin{equation}
\label{defNewton}
 p_{r}(x,y,k)= \sum_{i=1}^n
{x_{i}^r}+\frac{1}{k}\sum_{j=1}^m {y_{j}^r}
\end{equation}
belong to $\Lambda^0_{n,m;k}$ for all nonnegative integers $r$.


{\bf Theorem 2.} {\it If $k$ is not a positive rational number
then the algebra $\Lambda^0_{n,m; k}$ is generated by the deformed
Newton polynomials} $p_{r}(x,y,k), r \in {\bf Z}_+.$


Notice that for special values of $k$ this is not true. For
example, if $k = 1$ the deformed Newton sums generate the algebra
of symmetric polynomials in $n+m$ variables, while
$\Lambda^0_{n,m;k}$ is a much bigger algebra containing for
example $p = \prod (x_i - y_j)^3.$

To prove the Theorem let us recall that the {\it partition}
$\lambda$ of a natural number $N$ is a decreasing sequence of
non-negative integers $\lambda_1 \ge \lambda_2 \ge \lambda_3 \ge
...$ such that only a finite number of them are non-zero and their
sum is equal to $N$. This sum $\lambda_1 + \lambda_2 + \lambda_3 +
\dots $ is usually denoted as $|\lambda|.$ To each partition one
can relate a Young diagram with $N$ squares in a natural way (see
e.g. \cite{Ma}).


{\bf Proposition 2.} {\it If $k$ is not a positive rational then
the dimension of the homogeneous component $\Lambda^0_{n,m;k}$ of
degree $N$ is less or equal than the number of  partitions
$\lambda$ of $N$ such that $\lambda_{n+1}\le m$.}

\medskip

Notice that the corresponding Young diagrams are precisely the
ones contained in the {\it fat $(n,m)$-hook} (see Fig. 1).
\begin{center}

\includegraphics[width=7cm]{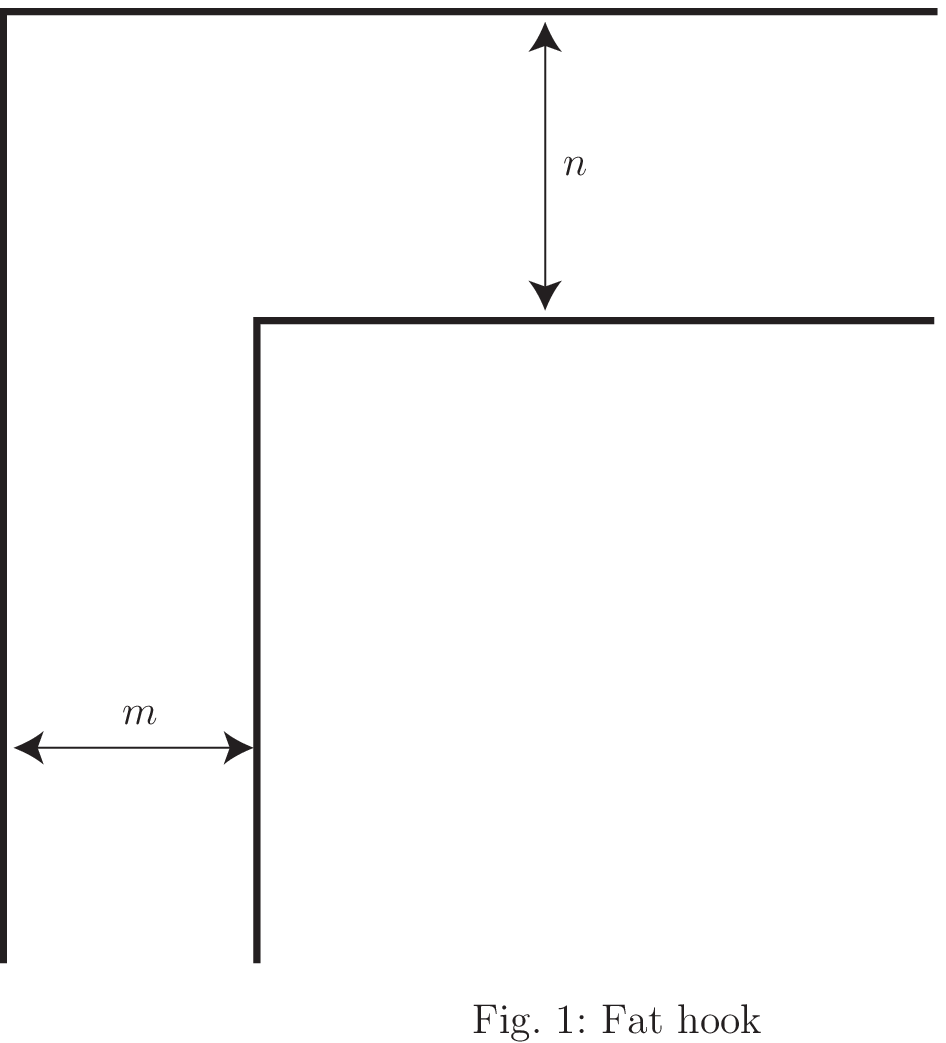}

\end{center}

Denote by $D_N(n,m)$  the number of such partitions (or diagrams
in this fat hook). Let $I=(i_{1},i_{2},\dots , i_{n})$ and
$J=(j_{1},j_{2},\dots , j_{m})$ be some  sequences (at the
beginning, unordered) of nonnegative integers such that
$\sum_{r=1}^n {i_{r}}+\sum_{s=1}^m {j_{s}}=N$. Let $N(J)$ be the
number of nonzero elements of $J$, $M(I,J)$ be the number of
elements of $I$ which are greater or equal to $N(J)$. Define the
sets $$ E_{reg}=\left\{ (I,J)\mid M(I,J)=n\right\}, \quad
E_{nreg}=\left\{ (I,J)\mid M(I,J)  <  n\right\}. $$ The pairs from
$E_{reg}$ will be called {\it regular}, otherwise - {\it
irregular}.

Let us prescribe to each pair $(I,J)$ a variable $C(I,J)$ in such
a way that $C(I,J) = C(\sigma(I), \tau(J)), \sigma \in S_n, \tau
\in S_m$ is the same for all orderings of $I$ and $J.$ It is easy
to see that the number of different $C(I,J), (I,J)\in E_{reg}$ is
equal to $D_N(n,m)$. For any sequence $I$ let us rewrite the
elements of $I$ in non-increasing order and denote this sequence
$I^+.$ Choose some integers $1\le r \le n$ and $1\le s\le m$ and
$1\le p\le N$ and consider the equations
\begin{equation}
\label{1} \sum_{i+j=p}(i-k j)C(I,J)=0
\end{equation}
where $i$ occupies the $r$-th place in  $I$ , $j$ occupies the
$s$-th place in $J$ and all other elements of $I,J$ are fixed. All
such equations form the system of linear equations on $C(I,J),$
which has the following meaning. Let $f$ be a homogeneous
polynomial of degree $N$, then the system (\ref{1}) is nothing
else but the condition $2)^0$ for the coefficients of $f \in
\Lambda^0_{n,m;k}$.

Thus to prove the Proposition it is enough to show that every
irregular $C(I,J), (I,J)\in E_{nreg}$ can be expressed from the
system (\ref{1}) as a linear combination of $C(I,J), (I,J)\in
E_{reg}$.

Let us first prove this statement for $n=1$.
We will use the induction with respect to the following total order on $E_{nreg}$. Let  $(I,J), (K,L)\in E_{nreg}$.
We will say that $(I,J) < (K,L)$ if for the corresponding ordered sets $(i,j_{1}^+\dots j_{m}^+$ and $(k,l_{1}^{+}\dots l_{m}^{+})$
either $N(J)< N(L)$ or $N(J)= N(L)$ and for $q=\min\{
i,k\}$ $(j_{m}^+,\dots ,j_{m-q}^+,i)< (l_{m}^{+},\dots,
l_{m-q}^{+},k)$ in the lexicographic order.

If $N(J)=1$ then we have $$k
jC(0,j)+(-1+(j-1)k)C(1,j-1)+(-2+(j-2)k)C(2,j-2)+\dots =0. $$ So we
have for $k \ne 0$ that $C(0,j)\in Span\{C(i,j),(i,j)\in
I_{reg}\}$

For general $N(J)>1$ take any $(I,J) \in  E_{nreg}$ with an
ordered $J=J^+$: $(i,j_{1},j_{2},\dots)$. Consider the equation
(\ref{1}) corresponding to $r=1, s=N(J)-i, p=i+j_{s}$. One can
check that $(I,J)$ is the largest pair with respect to our order
among all irregular pairs $(K,L)$ corresponding to $C(K,L)$
entering the equation with non-zero coefficients. Since the
coefficient at $C(I,J)$ is $-i +k j_s \neq 0$ (because $k$ is not
a positive rational) we can express $C(I,J)$ as a linear
combination of lower variables. This proves Proposition for $n=1$.

For general $n$ we can use the induction in $M(I,J)$. If
$M(I,J)=1$ we can use the previous arguments. Assume now that
$M(I,J) > 1.$ Consider one index $r,$ for which $i_{r} < N(J)$ and
apply previous arguments to express $C(I,J)$ as a linear
combination of $C(I^*,J^*)$ with $(I^*,J^*)$ such that $i_{r}^*\ge
N(J^*)$. According to inductive hypothesis we can express
$C(I^*,J^*)$ as a linear combination $C(I^{**},J^{**})$ where
$(I^{**},J^{**})\in I_{reg}$. Proposition 2 is proved.

Now let us prove the Theorem. Let us denote by ${\mathcal
N}_{n,m;k}$ the algebra generated by the deformed Newton sums
(\ref{defNewton}). As we have already mentioned ${\mathcal
N}_{n,m;k} \subset \Lambda^0_{n,m;k}.$ To show that ${\mathcal
N}_{n,m;k}= \Lambda^0_{n,m;k}$ it is enough to prove that the
dimension of the homogeneous component of degree $N$ of ${\mathcal
N}_{n,m;k}$ is not less than $D_{N}(n,m)$. To produce enough
independent polynomials we will use the theory of {\it Jack
polynomials} (see e.g. \cite{Ma}).

Let $\Lambda$ be algebra of symmetric functions in infinite number
of variables $z_{1},z_{2},\dots$ and
$p_{r}(z)=z_{1}^r+z_{2}^r+\dots$ be the power sum,
$P_{\lambda}(z,\theta)$ be Jack polynomial depending of partition
$\lambda$(see \cite{Ma} ). Consider a homomorphism $\phi$ from
$\Lambda$ to $\Lambda^0_{n,m;k}$ such, that $$ \phi (p_{r}(z))=
p_{r}(x,y,k). $$ Such a homomorphism was first used by Kerov,
Okounkov and Olshanski in \cite{KOO}. The image of the Jack
polynomials under this homomorphism sometimes is called super-Jack
polynomials (see e.g. \cite{Ok}).

One can show using some results from \cite{Ma} (see formulas
(7.9') and (10.19) from Chapter 6) that for $\theta = -k$
\begin{equation}
\label{Jack}
\phi (P_{\lambda}(z,\theta))=\sum_{\mu\subset\lambda} b_{{\lambda}
/{\mu}}(\theta) P_{\mu}(x,\theta)P_{{\lambda}^{'}
/{\mu}^{'}}(y,\theta^{-1})
\end{equation}
where $ b_{{\lambda}/{\mu}}$ is some rational function of $\theta$
with poles in non-positive rational numbers. Since $\theta = -k$
is not such a number by assumption these super-Jack polynomials
are well-defined.

From the formula (\ref{Jack}) it follows that the leading term in
lexicographic order of $\phi (P_{\lambda}(z,\theta))$ has a form
 $$
  x_{1}^{\lambda_{1}}\dots
x_{n}^{\lambda_{n}}y_{1}^{<{\lambda}^{\prime}_{1}-n>}\dots
  y_{m}^{<{\lambda}^{\prime}_{m}-n>}
 $$
where $\lambda^{\prime}$ is the partition conjugate to $\lambda$
and $<x>=\frac{x+|x|}{2} = max(0,x)$. From the definition $\phi
(P_{\lambda}(z,\theta))\in {\mathcal N}_{n,m;k}$. It is clear that
all these polynomials corresponding to the diagrams contained in
the fat hook are linearly independent. This completes the proof of
Theorem 2.

{\bf Remark.} The relation with the theory of Jack polynomials is
actually much deeper. We discuss this in detail in our paper
\cite{SV2} (see also \cite{Ser}, \cite{Ser1}).

As a corollary we can give a formula for Poincare series
$$P_{n,m}(t) = \oplus_i dim (\Lambda^0_{n,m;k})^{(i)}$$ of the
algebra $\Lambda^0_{n,m;k}$ for generic $k.$

{\bf Theorem 3.} {\it Poincare series of the algebra
$\Lambda^0_{n,m;k}$ for generic $k$ has the following form}
\begin{equation}
\label{Poincare}
P_{n,m}(t) =\frac{1}{(1-t)(1-t^2)\dots (1-t^n)}\left[1+\sum_{i=1}^m \frac{t^{i(n+1)}}{(1-t)(1-t^2)\dots (1-t^i)}\right].
\end{equation}

{\bf Proof.} From Theorem 2 it follows that the corresponding
Poincare series is the sum of $t^{|\lambda|}$ over all partitions
$\lambda$ which fit into fat $(n,m)$ hook: $$ P_{n,m}(t)=\sum_{
\lambda_{n+1}\le m} t^{|\lambda|}= \sum_{ \lambda_{n+1}=0}
t^{|\lambda|}+\sum_{ \lambda_{n+1}=1} t^{|\lambda|}+\dots+\sum_{
\lambda_{n+1}= m} t^{|\lambda|}, $$ where $|\lambda| = \lambda_1 +
\lambda_2 + ... +\lambda_N.$ It is easy to see that $$ \sum_{
\lambda_{n+1}= i}t^{|\lambda|} =\sum_{ \mu_{n+1}= 0} \sum_{
\nu_{i+1}= 0} t^{i(n+1)+|\mu|+|\nu|} =t^{i(n+1)}\sum_{ \mu_{n+1}=
0}t^{|\mu|}\sum_{ \nu_{i+1}= 0}t^{|\nu|}. $$ Since $$ \sum_{
\mu_{n+1}= 0}t^{|\mu|}=\frac{1}{(1-t)(1-t^2)\dots
(1-t^n)},\quad\sum_{ \nu_{i+1}=
0}t^{|\nu|}=\frac{1}{(1-t)(1-t^2)\dots (1-t^i)} $$ we arrive at
the formula (\ref{Poincare}).

{\bf Remark.}  Another (recurrent) formula for the generating function of the Young diagrams which fit into fat hook was found recently
by Orellana and Zabrocki \cite{OZ} in relation with the theory of the Schur functions and characters of Lie superalgebras.

Notice that the symmetry between $n$ and $m$ is not obvious from our formula (\ref{Poincare}) and leads to some identities which
might be interesting.

For $R = BC(n,m)$ the algebra $\Lambda^0_{R,B}$ is related to
$\Lambda^0_{n,m;k}$ in a very simple way: it is easy to check from
the definition that it consists of the polynomials $p(x_1^2,
x_2^2,..., y_1^2, y_2^2,..., y_m^2)$ where $p$ belong to
$\Lambda^0_{n,m;k}.$

{\bf Corollary.} {\it Poincare series $P^{BC}_{n,m}(t)$ of the
algebra $\Lambda^0_{R,B}$ for generalized system $R$ of type
$BC(n,m)$ and generic values of the deformation parameter has the
following form $$ P^{BC}_{n,m}(t) = P_{n,m}(t^2),$$ where
$P_{n,m}(t)$ is given by the formula} (\ref{Poincare}).

Let us discuss now the Harish-Chandra homomorphism. Let $R \in V$
be a generalized root system and $R^+$ be a set of positive roots.
Let us denote by $D[R^-]$ the algebra of differential operators on
$V^*$ with coefficients in ${\bf
C}[e^{-\alpha},(e^{-\alpha}-1)^{-1}],$ where $\alpha \in R^+$.

The {\it Harish-Chandra homomorphism} $\varphi :
D[R^-]\longrightarrow D,$ where $D$ is the algebra  of
differential operators on $V^*$ with constant coefficients, is
uniquely determined by the condition $\varphi (e^{-\alpha})=0$.
The algebra $D$ is isomorphic to the algebra of polynomial
functions on the space $V^*$.

Let now $R$ be a classical generalized root systems and consider
the algebra $Q_{R,m,B}$ generated by the quantum integrals $L_s$
of the corresponding deformed CMS problem (\ref{dCMS}): $$L_s =
{\hat\psi}_{0}\circ {\mathcal{L}}_s \circ {\hat\psi}_{0}^{-1}$$
where ${\hat\psi}_{0}$ is the multiplication operator by the
function ${\hat\psi}_{0}=\prod_{\alpha\in R_{+}}
\sin^{-m_{\alpha}}(\alpha,x)$ and ${\mathcal L}_s$ given by
(\ref{integrals}). It is easy to check that all the operators
$L_s$ belong to the algebra $D[R^-],$ so $Q_{R,m,B}$ is a
subalgebra in $D[R^-].$

{\bf Theorem 4.} {\it For generic values of the deformation
parameter the Harish-Chandra homomorphism maps the algebra of
quantum integrals of the deformed CMS problems $Q_{R,m,B}$ onto
the algebra $\Lambda_{R,B}$. In the rational limit the same is
true for the algebra} $\Lambda^0_{R,B}.$

{\bf Proof.} We will identify $V$ and $V^*$ using the form $B$.
Take $\lambda \in V$ and define
$x_{v}^{p}(\lambda)=e^{-(\lambda,x)}\varphi(\partial_{v}^{(p)})e^{(\lambda,x)}.$
From (\ref{recurrence}) we have the following recurrent relations
$$ x_{v}^{(p)}(\lambda)=(\lambda
,v)x_{v}^{(p-1)}(\lambda)-\frac{1}{2}\sum_{\alpha \in
R^+}m_{\alpha}(\alpha , v)\left
(x_{v}^{(p-1)}(\lambda)-x_{s_{<\alpha>}(v)}^{(p-1)}(\lambda)\right)
$$ which can be rewritten as $$ x_{v}^{(p)}(\lambda)=(\lambda-\rho
,v)x_{v}^{(p-1)}(\lambda ) +\frac{1}{2}\sum_{\alpha \in
R^+}m_{\alpha}(\alpha , v)x_{s_{<\alpha>}(v)}^{(p-1)}(\lambda).$$
The shifted functions
$y_{v}^{(p)}(\lambda)=x_{v}^{(p)}(\lambda+\rho)$ satisfy the
relations $$ y_{v}^{(p)}(\lambda)=(\lambda
,v)y_{v}^{(p-1)}(\lambda ) +\frac{1}{2}\sum_{\alpha \in
R^+}m_{\alpha}(\alpha , v)y_{s_{<\alpha>}(v)}^{(p-1)}(\lambda).$$
It is easy to see that the the image ${\mathcal Z}_{p}=\varphi
(L_{p})$ of the quantum integrals $L_{p}$ under Harish-Chandra
homomorphism has the form $$ {\mathcal Z}_{p}=\sum_{v\in\mathcal
O}\frac{y_{v}^{(p)}}{(v,v)}. $$ Let $(\lambda,\gamma)=0$, where
$\gamma\in R_{im}$ is an imaginary root.
 We should prove that
\begin{equation}
\label{Z} {\mathcal Z}_{p}\left
(\lambda-\frac{\gamma}{2}\right)={\mathcal Z}_{p}\left
(\lambda+\frac{\gamma}{2}\right ).
\end{equation}

We will prove this for the root system of type $A_{n,m}$, the case
of $BC_{n,m}$ root system is very similar.

 Without loss of generality we can assume that
$\gamma=e_{n}-e_{n+1}.$ Let us introduce
$y_{v}^{(p)-}(\lambda)=y_{v}^{(p)}(\lambda-\frac{\gamma}{2})$,
$y_{v}^{(p)+}(\lambda)=y_{v}^{(p)}(\lambda+\frac{\gamma}{2})$. We
have the following recurrent  relations $$
y_{v}^{(p)-}(\lambda)=\left(\lambda-\frac{1}{2}\gamma,v\right
)y_{v}^{(p-1)-}(\lambda ) +\frac{1}{2}\sum_{\alpha \in
R^+}m_{\alpha}(\alpha , v)y_{s_{<\alpha>}(v)}^{(p-1)-}(\lambda) $$
$$ y_{v}^{(p)+}(\lambda)=\left(\lambda+ \frac{1}{2}\gamma,v\right
)y_{v}^{(p-1)+}(\lambda ) +\frac{1}{2}\sum_{\alpha \in
R^+}m_{\alpha}(\alpha , v)y_{s_{<\alpha>}(v)}^{(p-1)+}(\lambda) $$
Let us denote by $v=e_n$, $u=s_{<\gamma>}(v)=e_{n+1}$ and
introduce $y_{v,u}^{(p)\pm}(\lambda)$ as $$
y_{v,u}^{(p)\pm}(\lambda)=((u,u)^{-1}+(v,v)^{-1})^{-1}[(u,u)^{-1}y_{u}^{(p)\pm}+(v,v)^{-1}y_{v}^{(p)\pm}].
$$

{\bf Lemma.} {\it On the hyperplane  $(\lambda,\gamma)=0$ the
following relations hold}:
\begin{equation} 1) y_{v,u}^{(p)\pm}(\lambda)
=(\lambda,v)y_{v,u}^{(p-1)\pm}(\lambda)+\frac{1}{2}\sum_{s_{<\alpha>}(v)\ne
u,v}m_{\alpha}{(\alpha,v)}y_{{s_{<\alpha>}(v)}}^{(p-1)\pm}(\lambda),
\label{rel1}
\end{equation}
2) $y_{v}^{(p)-}(\lambda) = y_{u}^{(p)-}(\lambda).$

Proof is by induction and based on the following fact, which can
be easily checked directly: if $s_{<\alpha>}(v)=s_{<\beta>}(u) $
then $$ ((u,u)^{-1}+(v,v)^{-1})^{-1}\left
[\frac{m_{\alpha}(\alpha,v)}{(v,v)}+\frac{m_{\beta}(\beta,u)}{(u,u)}\right]=m_{\alpha}(\alpha,v)=m_{\beta}(\beta,v).
$$

Now let $w = e_s$, $w\ne u,v$ then one can check that $$
(m_{\alpha}(\alpha,w)y_{v}^{(p-1)+}+m_{\beta}(\beta,w)y_{u}^{(p-1)+}=(m_{\alpha}(\alpha,w)+m_{\beta}(\beta,w))y_{v,u}^{(p-1)+}
$$ where $ s_{<\alpha>}(v)=s_{<\beta>}(u)=w $. Notice that the
last relation determines $\alpha$ and $\beta$ uniquely in our
case. Using this one can show that
\begin{equation}
\label{rel2}
 y_{w}^{(p)\pm}
=(\lambda,w)y_{w}^{(p-1)\pm}+\frac{1}{2}(m_{\alpha}(\alpha,w)+m_{\beta}(\beta,w))y_{v,u}^{(p-1)\pm}+
\frac{1}{2}\sum_{s_{<\delta}>(w)\ne u,v } m_{\delta}(\delta,w)
y_{s_{<\delta>(w)}}^{(p-1)\pm}
\end{equation}
provided $(\lambda,\gamma) =0.$

From the relations (\ref{rel1}), (\ref{rel2}) it follows that
$y_{w}^{(p)+}=y_{w}^{(p)-}, y_{v,u}^{(p)-}=y_{v,u}^{(p)+}$ on the
hyperplane $(\lambda,\gamma)=0$, which imply the relation
(\ref{Z}).

Note that the highest term of ${\mathcal Z}_{r} (\lambda)$ is
\begin{equation}
\label{summa}
\lambda_{1}^r+\dots+\lambda_{n}^r+k^{r-1}(\lambda_{n+1}^r+\dots+\lambda_{n+m}^r).
\end{equation}
Now from Theorem 2 it follows that for generic $k$ homomorphism
$\varphi$ is surjective. The fact that it is injective is obvious.
This completes the proof in trigonometric case, rational case
easily follows. In fact $\omega$ in the definition of
$\Lambda^{\omega}_{R,B}$ and $\omega$ in the limiting procedure
from trigonometric to rational case could be identified.

{\bf Remark.} Notice that we have proved that the image of
Harish-Chandra homomorphism belong to the algebra $\Lambda_{R,B}$
for {\it all} values of the parameter $k$. The condition that $k$
is generic is used only to claim that the image coincides with
this algebra.

As a corollary we have the following statement which is probably
true for all generalized root systems and all values of
deformation parameter.

{\bf Proposition 3.} {\it For the classical generalized root
systems and generic values of the deformation parameter the
algebra $\Lambda^{0}_{R,B}$ is the associated graded algebra for}
$\Lambda_{R,B}.$

In $A(n,m)$ case we can give an explicit formula for the
generators of the algebra $\Lambda_{R,B}:$
\begin{equation}
\label{generators} Y_r(\lambda) = \sum_{i=1}^{n} B_r(\lambda_i +
1/2) + k^{r-1} \sum_{j=1}^m B_r(\lambda_{j+n} +1/2),
\end{equation}
where $B_r(x)$ are the classical Bernoulli polynomials. One can
easily check using the relation $B_r(x+1) - B_r(x) = rx^{r-1}$
that $Y_r$ satisfy the relations (\ref{Z}) and have the highest
term (\ref{summa}).

 We finish this section with the following

{\bf Theorem 5.} {\it Algebra $\Lambda^0_{n,m;k}$ is finitely
generated if and only if $k$ is not a negative rational number of
the form $-\frac{s}{r},$ where $1 \le r \le n, 1 \le s \le m$.}

Consider the subalgebra $P(k) = {\bf C}[p_1, ...p_{n+m}]$
generated by the first $n+m$ deformed Newton sums
(\ref{defNewton}). We need the following result about common zeros
of these polynomials (cf. Proposition 1 in \cite{FV2}).

{\bf Proposition 4.} {\it Consider the following system of
algebraic equations $$ \left\{
\begin{array}{rcl}
x_{1}+x_{2}+\dots+x_{n}+k^{-1}(x_{n+1}+x_{n+2}+\dots+x_{n+m})=0\\
x_{1}^2+x_{2}^2+\dots+x_{n}^2+k^{-1}(x_{n+1}^2+x_{n+2}^2+\dots+x_{n+m}^2)=0\\
\cdots\\
x_{1}^{n+m}+x_{2}^{n+m}+\dots+x_{n}^{n+m}+k^{-1}(x_{n+1}^{n+m}+x_{n+2}^{n+m}+\dots+x_{n+m}^{n+m})=0\\
\end{array}
\right.$$ If parameter $k$ is not a negative rational number of
the form $-\frac{s}{r},$ where $1 \le r \le n, 1 \le s \le m$ then
the system has only trivial (zero) solution in ${\bf C}^{n+m}$.
Converse statement is also true.}

To prove this suppose that the system has a nontrivial solution
$x_{1},\dots,x_{n+m}$. We can assume that $x_{i}\ne 0$ for all
$i=1,\dots, n+m$. Let us re-group the set $X =\{
x_{1},x_{2},\dots,x_{n+m}\} \subset {\bf C}$ identifying equal
$x_i$'s as $\{z_{1}, \dots, z_{p}\}, p \le n+m, $ where all $z_j$
are different. Multiplicity of $z_j$ is a pair $(r_j, s_j)$, where
$r_j$ shows how many times $z_j$ enters the set $\{
x_{1},x_{2},\dots,x_{n}\}$ and $s_j$ is the same for the rest of
the set $X$. For the numbers $z_j,\quad 1 \le j \le p$ we have the
system $$\sum_{j=1}^{p} a_j z_j^i =0, \quad i=1,\dots, n+m,$$
where $a_j = r_j + k^{-1} s_j.$ Consider the first $p$ of these
equations as the linear system on $a_j.$ Its determinant is of
Vandermonde type and is not zero since all $z_j$ are different and
non-zero. Hence all $a_j$ must be zero which may happen only if $k
= -\frac{s}{r}$ for some $1 \le r \le n, 1 \le s \le m.$ The
converse statement is obvious: if $k = -\frac{s}{r}$ then we can
take $x_1 = x_2 = \dots = x_r = z = x_{n+1} = \dots = x_{n+s}$ and
other $x_i$ being zero to have the non-trivial solutions of the
system with arbitrary $z$.

From Proposition 4 it follows that for $k \neq -\frac{s}{r}$ the
algebra of all polynomials on $V$ is a finitely generated module
over subalgebra $P(k).$ By a general result from commutative
algebra (see e.g. Proposition 7.8 from \cite{Atiyah-Macdonald})
this implies that $\Lambda^0_{n,m;k}$ is finitely generated.

Now suppose that $k = -\frac{s}{r}$ for some $1 \le r \le n, 1 \le
s \le m.$ Consider the following homomorphism $$\phi_{r,s} :
\Lambda^0_{n,m;k} \rightarrow \Lambda^0_{1,1; -1}$$ by sending a
polynomial $f (x_{1},\dots,x_{n},y_{1},\dots,y_{m})$ into $$\hat
f(x,y) = f(x, x, \dots, x, 0, \dots, 0, y, y, \dots, y, 0, \dots,
0),$$ where $x$ is repeated $r$ times and $y$ is repeated $s$
times. One can easily check that if $f$ is in $\Lambda^0_{n,m;k}$
with this particular $k$ then $\hat f(x,y)$ satisfies the
condition $(\partial_x + \partial_y) \hat f = 0$ when $x=y$, i.e.
$\hat f$ belongs to $\Lambda^0_{1,1; -1}.$ The last algebra is
actually very simple: it consists of the polynomials which are
constant on the line $x=y$ and thus have the form $c +
q(x,y)(x-y)$ with arbitrary polynomial $q.$ It is easy to see that
this algebra is not finitely generated. Since it is a homomorphic
image of the algebra $\Lambda^0_{n,m;k}$ this implies that the
algebra $\Lambda^0_{n,m;k}$ with $k = -\frac{s}{r}$ is also not
finitely generated. Theorem 5 is proved.

{\bf Remark.} The case $k =-1$ is actually very special: in that
case the algebra $\Lambda^0_{n,m;k}$ is known as algebra of {\it
supersymmetric polynomials} and plays an important role in
geometry (see e.g. \cite{FP}, Chapter 3). Notice also that the
special values of $k$ in the theorem 2 are positive rationals
while in theorem 5 they are negative.

{\bf Corollary.} {\it For generic values of the deformation
parameters the algebras $\Lambda^0_{R,B}$ and $\Lambda_{R,B}$ for
the classical generalized root systems are finitely generated.}

Indeed, from Proposition 3 it follows that it is enough to show
that $\Lambda^0_{R,B}$ is finitely generated. For $A(n,m)$ type
this follows directly from Theorem 5, to prove this for $BC(n,m)$
one should replace in Theorem 5 all the coordinates by their
squares.

An interesting question is whether the algebra $\Lambda^0_{n,m;k}$
is free as a module over its polynomial subalgebra $P(k)$, i.e.
has Cohen-Macaulay property. If this is true (which we believe to
be so) then our formula (\ref{Poincare}) gives the degrees of its
generators for generic values of $k.$

For example, when $m=1$ we have $$P_{n,1}(t)
=\frac{1}{(1-t)(1-t^2)\dots (1-t^n)}\left[1+
\frac{t^{n+1}}{(1-t)}\right] = \frac{1 + t^{n+2} + t^{n+3} + \dots
+ t^{2n+1}}{(1-t)(1-t^2)\dots (1-t^{n+1})},$$ which shows that the
generators should have the degrees $0, n+2, n+3,..., 2n+1.$ The
conjecture is that one can take the corresponding deformed Newton
sums as such generators.

For $n=2$ this is in a good agreement with the results from
\cite{FV2}, where the Cohen-Macaulay property for the rings of
quasi-invariants related to $A(n,1)$ and $BC(n,1)$ is established
(for any $n$) and the corresponding Poincare series are found (for
$n=2$).

\section{Generalizations: elliptic and difference versions.}

The deformed quantum CMS systems we discussed have some natural
generalizations. First of all if we replace in all the formulas
for these operators the function $\frac{1}{\sin^2 z}$ by
Weierstrass' elliptic function $\wp(z)$ we will have the {\it
deformed elliptic CMS operators}.

For the generalized root system of type $A(n-1,m-1)$ we have

\begin{eqnarray}
\label{eanm} L^{ell}_{A(n-1,m-1)}&=&-\left
(\frac{\partial^2}{{\partial x_{1}}^2}+\dots
+\frac{\partial^2}{{\partial x_{n}}^2}\right)-
k\left(\frac{\partial^2}{{\partial y_{1}}^2}+\dots
+\frac{\partial^2}{{\partial y_{m}}^2}\right) \nonumber\\& &
+\sum_{i<j}^{n}
2k(k+1)\wp(x_{i}-x_{j})+\sum_{i<j}^{m}2(k^{-1}+1)\wp(y_{i}-y_{j})
\nonumber \\& &
 +\sum_{i=1}^{n}\sum_{j=1}^{m} 2(k+1)\wp(x_{i}-y_{j}).
\end{eqnarray}

For $BC(n,m)$ we can write a more general {\it deformed Inozemtsev operator}:
\begin{eqnarray}
\label{ebcnm}
L^{ell}_{BC(n,m)}&=&-\left(\frac{\partial^2}{{\partial
x_{1}}^2}+\dots +\frac{\partial^2}{{\partial x_{n}}^2}\right)
-k\left(\frac{\partial^2}{{\partial y_{1}}^2}+ \dots
+\frac{\partial^2}{{\partial y_{m}}^2} \right) \nonumber \\& &
+\sum_{i<j}^{n}2k(k+1)(\wp(x_{i}-x_{j})+\wp(x_{i}+x_{j}))
+\sum_{i<j}^{m}2(k^{-1}+1)(\wp(y_{i}-y_{j})+\wp(y_{i}+y_{j}))\nonumber
\\& &
+\sum_{i=1}^{n}\sum_{j=1}^{m}2(k+1)(\wp(x_{i}-y_{j})+\wp(x_{i}+y_{j}))
\nonumber
\\& & +\sum_{i=1}^n \sum_{l=0}^{3} q_l(q_l+1) \wp
(x_{i}+\omega_l)+ \sum_{j=1}^m \sum_{l=0}^{3} s_l(s_l+1) \wp
(y_{j}+\omega_l)
\end{eqnarray}
where $\omega_0 = 0, \quad \omega_l, l=1,2,3$ are the half-periods
of the corresponding elliptic curve and the 9 parameters
$k,q_l,s_l, l=0,1,2,3 $ satisfy the following 4 relations
\begin{equation}
\label{erel}
\quad 2q_l+1=k(2s_l+1)
\end{equation}
for all $l$.


We conjecture that the elliptic versions of the deformed CMS
problems are integrable as well. For $m=1$ some results in this
direction are found in \cite{Kh},\cite{KhP} (see also recent paper
\cite{CEO}).

Consider now the difference case.  Let us introduce the following
{\it deformed Macdonald-Ruijsenaars operator.} It depends on two
parameters $t$ and $q$ and has the form
\begin{equation}
\label{defMR} D^{n,m}=\frac{1}{1-q}\sum_{i=1}^n
A_{i}T_{q,x_{i}}+\frac{1}{1-t}\sum_{j=1}^m B_{j}T_{t,y_{j}}
\end{equation}
where
$$ A_{i}=\prod_{k\ne i}^{n}
\frac{(x_{i}-tx_{k})}{(x_{i}-x_{k})}\prod_{j=1}^{m}
\frac{(x_{i}-qy_{j})}{(x_{i}-y_{j})}, \quad B_{j}=\prod_{i=1}^{n}
\frac{(y_{j}-tx_{i})}{(y_{j}-x_{i})}\prod_{l\ne j}^{m}
\frac{(y_{j}-qy_{l})}{(y_{j}-y_{l})}
$$
and
$T_{q,x_{i}},T_{t,y_{j}}$ are the "shift operators":
$$(T_{q,x_{i}}f)(x_{1},\dots,x_{i},\dots,x_{n},y_{1},\dots,y_{m})=f(x_{1},\dots,qx_{i},\dots,x_{n},y_{1},\dots,y_{m})
$$ $$
(T_{t,y_{j}}f)(x_{1},\dots,x_{n},y_{1},\dots,y_{j},\dots,y_{m})=f(x_{1},\dots,x_{n},y_{1},\dots,ty_{j},\dots,y_{m}).
$$ For $m=1$ a similar operator was considered by O. Chalykh in
\cite{Ch2}. We should mention that there is a small discrepancy
between his and our formulas because of the misprint in the
formula (7.3) in \cite{Ch2}.

It is interesting that the form of the operator (\ref{defMR}) is
invariant under the simultaneous interchange of $q \leftrightarrow
t$ and $x \leftrightarrow y.$ In a way the deformed operator
(\ref{defMR}) is more symmetric than the original
Macdonald-Ruijsenaars operator \cite{Ma}, \cite{R}. In the
differential case this duality corresponds to the invariance of
the family of the deformed CM operators (\ref{anm}) under the
interchange $x \leftrightarrow y$ and $k \leftrightarrow
\frac{1}{k}.$

One can verify that the operator (\ref{defMR}) can be rewritten in
terms of the root system $R$ of type $A(n,m)$ as follows
(notations are as in Section 3)
\begin{equation}
\label{defM} D^{R}=\sum_{v\in \mathcal O}
\frac{1}{1-q^{(v,v)}}\left[\prod_{\alpha\in R, <\alpha,v>>0}
\frac{1-t_{\alpha}^{(\alpha,v)}q^{-\alpha}}{1-q^{-\alpha}}\right ]
T_{v},
\end{equation}
where $$t_{\alpha} = q^{m_{\alpha}}, \quad
q^{\alpha}(u)=q^{(\alpha,u)}, \quad (T_{v}f)(u)=f(u+v)$$ and the
relation with the previous formula is given by $$t = q^k, \quad
x_i = q^{e_i},  \quad y_j = q^{e_{n+j}}.$$

In this form the operator can be immediately generalized for any
reduced generalized root system $B_{n,m}, C_{n,m}, D_{n,m}$ (but
not for the general $BC(n,m)$ case). We intend to discuss the
properties of these deformed Macdonald operators in a separate
paper.

\section*{Concluding remarks.}

We have constructed for each generalized root system a family of
the deformed quantum CMS problems and proved their integrability
for the classical series. The proof is effective but not
conceptual. We present more conceptual proof for $A(n,m)$ series
in our paper \cite{SV2}, where the algebraic varieties
corresponding to the rings $\Lambda^0_{R,B}$ and relations with
the theory of Jack polynomials are also discussed in more detail.

It is clear that these relations between the theory of Lie
superalgebras and quantum integrable systems should be understood
better. In particular, the spectral theory for the deformed CMS
operators should be related to the representation theory of Lie
superalgebras and spherical functions on the symmetric
superspaces. We hope to come back to this problem soon.

\section*{Acknowledgements.}
We are grateful to O. Chalykh, A. Okounkov, G. Olshanski and
especially to M. Feigin for useful discussions and helpful
remarks.

This work was supported by EPSRC (grants GR/R70194/01 and
GR/M69548). The second author (A.P.V.) is grateful to IHES
(Bures-sur-Yvette, France) for the hospitality in February 2003
when the final version of this paper was prepared.

\end{document}